\documentclass[12pt,a4paper]{article}
\usepackage{longtable} %allows multipage tables
\usepackage{ragged2e} % ragged right modifier useful in tables
\usepackage{array} %for column types below 
\newcolumntype{C}[1]{>{\centering\arraybackslash}p{#1}}
\newcolumntype{P}[1]{>{\raggedright\arraybackslash}p{#1}}
\usepackage[table]{xcolor} % coloured cells in tables http://ctan.org/pkg/xcolor
\usepackage{multirow} %allows cells to expand over multi. rows
\usepackage{multicol} %allows cell to expand over multi. columns
\definecolor{table-1}{RGB}{252,252,252} 

\usepackage{listings}

\usepackage{amsthm}

\usepackage{amsmath}

\usepackage{enumitem}

\usepackage{amsmath}

\usepackage{afterpage}

\usepackage{graphicx} %allow you to include figures
\usepackage[font=small,labelfont=bf]{caption}
\usepackage[font=small,labelfont=bf]{subcaption}
\usepackage{wrapfig}
\usepackage{caption}
\usepackage{adjustbox}
\usepackage[justification=centering]{caption} %centres captions

\usepackage{alphalph} %allows numbering to extend beyond 26

\usepackage[left=2cm,right=2cm,top=2cm,bottom=2.2cm,headsep=18pt]{geometry} %page margins
\usepackage{pdflscape} %allows landscape pages
\usepackage{float}
\usepackage{fancyhdr}
\usepackage{titlesec} %controls title spacing
\usepackage{color} 
\titlespacing*{\section}{0pt}{.2cm plus .2cm}{-.1cm plus .1cm}
\titlespacing*{\subsection}{0pt}{.2cm plus .1cm}{-.2cm plus .1cm}
\titlespacing*{\subsubsection}{0pt}{.1cm plus .1cm}{-.3cm plus .1cm}
\usepackage{amsmath}
\usepackage{amsfonts}
\usepackage{amssymb}
\usepackage{bm} %bold math
\usepackage{relsize}

\usepackage[utf8]{inputenc}
\usepackage[english]{babel}

\usepackage{fancyhdr}

\usepackage[colorlinks,citecolor=black,linkcolor=black,urlcolor=black,bookmarks=false,hypertexnames=true]{hyperref} %allows citations and figure numbers and table of contents etc. to be hyper linked 
\usepackage{cite}
\usepackage{url}
\usepackage{setspace} %change spacing in references
\usepackage{cleveref} %allows smart referencing of multiple equations

\makeatletter
\newcommand\ackname{Acknowledgements}
   \makeatother
\usepackage[T1]{fontenc}
\usepackage{soul}
\usepackage{sectsty}
\usepackage{lscape}
\usepackage{comment}
\usepackage{amsmath}

\begin{document}

{\fontfamily{cmr}\selectfont
%This text uses a different font typeface
}

\setcounter{page}{2}
\fancyhf{}
\renewcommand{\headrulewidth}{0pt}
\fancyfoot[C]{\thepage}

%Title page
\title{\line(1,0){500}\\\textsc{Handling Overlapping Asymmetric Datasets - A Twice Penalized P-Spline Approach}\\\line(1,0){500}}

\author{\small
    \textbf{Matthew McTeer$^1$,\; Robin Henderson$^2$,\; Quentin M. Anstee$^3$,\; Paolo Missier$^1$} \\
    \scriptsize
    $^1$ School of Computing, Faculty of Science, Agriculture \& Engineering, Newcastle University \\
    \scriptsize
    $^2$ School of Mathematics, Statistics and Physics, Faculty of Science, Agriculture \& Engineering, Newcastle University \\
    \scriptsize
    $^3$ Translational \& Clinical Research Institute, Faculty of Medical Sciences, Newcastle University\\
}
\date{17\textsuperscript{th} November 2023} \maketitle  \thispagestyle{empty}

%Abstract
\begin{changemargin}{1.5cm}{1.5cm} 
\begin{center}
  \normalsize\bfseries{\textsc{Abstract}}
\end{center}
\footnotesize Overlapping asymmetric datasets are common in data science and pose questions of how they can be incorporated together into a predictive analysis. In healthcare datasets there is often a small amount of information that is available for a larger number of patients such as an electronic health record, however a small number of patients may have had extensive further testing. Common solutions such as missing data imputation can often be unwise if the smaller cohort is significantly different in scale to the larger sample, therefore the aim of this research is to develop a new method which can model the smaller cohort against a particular response, whilst considering the larger cohort also. Motivated by non-parametric models, and specifically flexible smoothing techniques via generalized additive models, we model a twice penalized P-Spline approximation method to firstly prevent over/under-fitting of the smaller cohort and secondly to consider the larger cohort. This second penalty is created through looking at discrepancies in the marginal value of covariates that exist in both the smaller and larger cohorts. Through a series of data simulations, penalty parameter tunings and model adaptations to consider both a continuous and binary response, we find that our twice penalized approach offers an enhanced model fit over a linear B-Spline model and once penalized P-Spline approximation method. Applying our twice penalized method to a real-life healthcare dataset relating to an individual’s risk of developing Non-Alcoholic Steatohepatitis, we see an improved model fit performance of over 65\% as opposed to linear and once penalized methods. Areas for future work within this space include adapting our method to not require dimensionality reduction and also consider parametric modelling methods. However, to our knowledge this is the first work to propose additional marginal penalties in a flexible regression of which we can report a vastly improved model fit that is able to consider asymmetric datasets, without the need for missing data imputation.  
\end{changemargin}

\newpage
\pagenumbering{arabic}
%table of contents
%\begin{spacing}{1.15} %change this number to get table of contents onto 1 page
%\setcounter{tocdepth}{3}
%setcounter{secnumdepth}{3}
%\tableofcontents
%\addtocontents{toc}{\vspace{-3.5mm}}
%reduces spacing below contents title 
%\end{spacing}
%\newpage

%\pagestyle{fancy}
\fancyhf{} %sets all head and foot elements empty
\renewcommand{\sectionmark}[1]{\markboth{\thesection\ #1}{}}
\renewcommand{\subsectionmark}[1]{\markright{\thesubsection\ #1}}
%\fancyhead[L]{\small\leftmark}
%\fancyhead[R]{\small\rightmark}
%\fancyfoot[C]{\thepage}

%Main Text
% \title{Paper 2}  % Commented by Nav

\section{\textsc{Purpose - Horizontal and Vertical Data}}

It is a common problem in data science and statistics that for a small number of observations there is a greater level of information that is known, which becomes difficult to incorporate into analysis utilising a larger cohort of observations of which far less information is understood. In the example of healthcare studies, there may be a large number of individuals of which their basic information is known, such as an electronic health record, however a select number of individuals may have had further testing such as having their genetic information recorded. This problem can also translate into how to deal with large levels of missing data. Kang (2013) presents several techniques for handling missing data, including case deletion, mean substitution and multiple imputation \cite{kang2013prevention}. Indeed one of the most common and popular tools for handling missing data is missing imputation tool MICE (Multiple Imputation by Chained Equations), which is an available library in a vast number of coding languages \cite{van2011mice}. However when there is a large proportion of missing data, Multiple Imputation (MI) is not considered to be the most effective way of dealing with missing data issues \cite{lee2021evaluation}. Many authors have attempted to provide 'cutoff' points for an acceptable amount of missing data that MI can handle \cite{schafer1999multiple, bennett2001can} however these are found to be largely arbitrary and other factors need to be taken into account such as types of missingness and imputation mechanisms. It is however considered that vast amounts of missing data are not suited to MI and other methods are necessary. This paper explores how we can utilise the larger cohort of individuals to enhance what is learnt from the smaller cohort, without the need for imputation.

Let us assume that there are two datasets: 
\setlist{nolistsep}
\begin{itemize}
\itemsep0em
    \item Horizontal Data, denoted $\mathcal{H}$
    \item Vertical Data, denoted $\mathcal{V}$
\end{itemize}

For simplicity, $\mathcal{H}$ has a sample size of $N_\mathcal{H}$ with two scalar covariates $x$ and $z$, $\mathcal{V}$ has a sample size of $N_\mathcal{V}$ with only one covariate $x$ and in our case $N_\mathcal{V}$ $\gg$ $N_\mathcal{H}$. Both datasets contain a response variable $y$. In reality, $x$ and $z$ would represent a selection of covariates each as we will show in the Application section of this work. The validity of reducing multiple covariates into single $x$ and $z$ vectors is discussed later. Illustrated in Figures \ref{fig:fig1} and \ref{fig:fig2} is how each datasets may look in practise: 

\begin{figure}[H]
    \centering
    \includegraphics[width=0.4\textwidth]{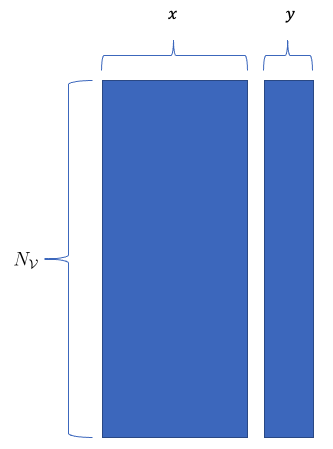}
    \caption{Vertical Data, $\mathcal{V}$}
    \label{fig:fig1}
\end{figure}
\vspace{10mm}

\begin{figure}[H]
    \centering
    \includegraphics[width=0.7\textwidth]{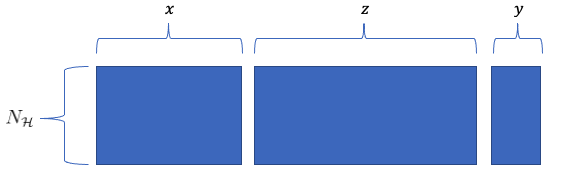}
    \caption{Horizontal Data, $\mathcal{H}$}
    \label{fig:fig2}
\end{figure}

From these diagrams we see $\mathcal{V}$ dataset is tall and thin, and $\mathcal{H}$ dataset is short and wide, hence the naming, Vertical and Horizontal datasets. 

The motivation for this work was that in previous experiments we had modelled using a set of easily accessible covariates obtained from routine GP appointments upon a series of binary responses relating an individual’s risk of various endpoints of Non-Alcoholic Fatty Liver Disease (NAFLD), ranging from relatively benign steatosis to more aggressive fibrosis and cirrhosis. In the context for this paper, these covariates would be considered as $x$ and the modelling that was undertaken would have been upon $\mathcal{V}$ only. For a small number of the individuals within $\mathcal{V}$ we now have their genomic sequencing data available, which for the purposes of this paper we consider to be $z$. We now wish to incorporate $z$ into the analysis of being able to predict binary NAFLD endpoints, however with $N_\mathcal{H}$ being far smaller than $N_\mathcal{V}$, this is not a simple task. This provides a more general research area of whether it is possible to utilise what we learn from $\mathcal{V}$ to enhance the predictive performance and modelling upon observations within $\mathcal{H}$, thus improving our knowledge about response variable $y$. 

By firstly displaying a knowledge and motivation for studying non-parametric modelling techniques, we specifically focus upon smoothing methods and penalized regression modelling including B-Splines and P-Splines. We show that they display specific qualities that make them attractive for modelling our smaller cohort $\mathcal{H}$ and also show that they can be adapted into a new model that is able to consider the larger cohort $\mathcal{V}$ through including a second penalty term which takes into account discrepancies in the marginal value of $x$, i.e. covariates that exist in both $\mathcal{H}$ and $\mathcal{V}$. We compare our twice penalized model structure against a linear B-Spline model and single penalty P-Spline estimation upon a series of controlled data simulations, before adapting our model further to take into account a binary response $y$. We will finally apply our model upon our motivating real data example, specifically using easily accessible covariates obtained from routine GP appointments alongside genomic sequencing data of which fewer individuals have, to predict an individual's personal risk of developing Non-Alcoholic Steatohepatitis (NASH).

\section{\textsc{Background}}
Broadly speaking there are two branches of statistical modelling that exist: parametric statistics and non-parametric statistics. Parametric modelling assumes that data can be modelled appropriately through a specific probability distribution with certain parameters, whereas non-parametric modelling does not assume that data follows a particular distribution. Within this section we will outline first of all our motivation for using non-parametric models over parametric models, before going into detail about the non-parametric techniques we use throughout this paper. 

\subsection{\textsc{Motivation for Non-Parametric Models}}
Considering a response $y$ and two vector covariates $x$ and $z$, we denote $f(.)$ as a generic notation for probability functions, whether for discrete or continuous random variables. We are particularly interested in the conditional probability functions $f(y|x)$ and $f(y|x,z)$, with the former being the marginal after integrating out $z$ of the latter. In general the relationship is:
\begin{equation*}
    f(y|x) = \int f(y,z|x) dz = \int f(y|x,z)f(z|x) dz
\end{equation*}
When determining the suitability of either a parametric or non-parametric approach to modelling both probability functions, we note that if $f(y|x,z)$ takes a parametric modelling form, it is not usually possible for $f(y|x)$ to be the same parametric form also.

\underline{\textit{Example:}} Assume $y$ is binary and suppose the full conditional is logistic:
\begin{equation*}
    f(y|x,z)=Pr(y=1|x,z) = \text{expit}(\beta_0+\beta_1x+\beta_2z)
\end{equation*}
where $\text{expit}(a)=e^{a}/(1+e^{a})$.
Take $z$ to be a binary scalar that is independent of $x$ with $Pr(z=1)=1/2$, then:
\begin{equation*}    
\begin{split}
    f(y|x,z=0) &= \text{expit}(\beta_0+\beta_1x) \\
    f(y|x,z=1) &= \text{expit}(\beta_0+\beta_1x+\beta_2) \\
    \\
\end{split}
\end{equation*}
Therefore:
\begin{equation*}
\begin{split}
    f(y|x) &= \frac{1}{2}\biggl(\text{expit}(\beta_0+\beta_1x)+\text{expit}(\beta_0+\beta_1x+\beta_2)\biggr)\\
    &\neq \text{expit}(\beta_0+\beta_1x)
\end{split}
\end{equation*}

Hence $f(y|x)$ is not of logistic form. We are therefore motivated to look at non-parametric models to take into account conditional models.

\subsection{\textsc{Flexible Smoothing with Splines}}
Many non-parametric modelling techniques exist; one popular approach is smoothing, in particular spline methods. Eilers and Marx \cite{eilers1996flexible} cite several reasons for their popularity including datasets being too complex to be modelled sufficiently through parametric models, and also an increasing demand for graphical representations and exploratory data analysis. We will first of all demonstrate what is meant by the term 'smoothing' before introducing spline functions as a common approach, highlighting their key properties and popular kinds of spline that exist. We will then discuss two specific kinds of spline in detail, B-Splines and P-Splines, which form the basis of our research throughout this work.  

\subsubsection{\textsc{An Introduction to Smooth Functions}}
Let us assume that $x$ is a vector. A linear model therefore assumes:
\begin{equation*}
    E[y|x] = \beta_0 + \beta_1x_1 + ...
\end{equation*}
A generalised linear model assumes that:
\begin{equation*}
    g(E[y|x]) = \beta_0 + \beta_1x_1 + ...
\end{equation*}
where $g(.)$ is some function.

Motivated by non-parametric classes of model, we introduce generalised additive models (GAMs), first proposed by Hastie and Tibshirani (1986) \cite{hastie1990generalized}. GAMs build upon familiar likelihood-based regression models in a way that provide more robustness and flexibility, such that more complex distributed data points can be modelled beyond linear or polynomial regression. If there is a single covariate, a GAM assumes a model that is: 
\begin{equation*} 
    g(E[y|x]) = \beta_0 + \gamma(x)
\end{equation*}
where $\gamma(.)$ is a smooth function. With regards to how we select $x$, one way is through a polynomial model that is of the form:
\begin{equation*} 
    g(E[y|x]) = \beta_0 + \beta_1x + \beta_2x^2 + \beta_3x^3 + ...
\end{equation*}
However, a more flexible alternative is provided by the use of basis functions:
\begin{equation*} 
    g(E[y|x]) = \beta_0 + \beta_1S_1(x)+\beta_2S_2(x) + \beta_3S_3(x) + ...
\end{equation*}
In this case, $S_1(.),S_2(.),S_3(.)$ etc are smooth basis functions, and we can receive our polynomial model above by letting $S_1(x)=x$, $S_2(x)=x^2$, $S_3(x)=x^3$ and so on; these smooth basis functions can be displayed within a basis matrix, with each row being evaluated at different values for $x$.

\subsubsection{\textsc{An Introduction to Splines}}

Common basis functions are spline basis functions. Spline models split the  $x$-axis into separate intervals and assume a different model for each, for example:
\begin{equation*}
        S_1(x) =
        \begin{cases}
            S_1^*(x), & 0<x\leq1\\
            
            0, & \text{otherwise.}
        \end{cases}
\end{equation*}

\begin{equation*}
        S_2(x) =
        \begin{cases}
            S_2^*(x), & 1<x\leq2\\
            
            0, & \text{otherwise.}
        \end{cases}
\end{equation*}
The joins between each interval are known as 'knots'. In order for the function to be differentiable everywhere and therefore smooth at the knots, the following conditions must also hold:
\begin{equation*}
    S_1(1) = S_2(1) , \hspace{10 mm} S_1'(1) = S_2'(1), \hspace{10 mm} S_1''(1)=S_2''(1), ...
\end{equation*}

In practice it is usually sufficient that the derivatives up to $S''(x)$ match at the knots, this is because the human eye struggles to detect higher order discontinuities \cite{perperoglou2019review}. The number and placement of knots, along with the choice of smooth polynomial pieces that are fitted between two consecutive knots is what defines the type of spline. This along with the use or not of a penalty function are the key choices that need to be made when using splines; we will talk about the inclusion of a penalty term within spline modelling later, however for now we focus upon non-penalized splines known as 'regression splines'.

There are many kinds of regression splines that exist, all of which differ with regards to their statistical properties. De Boor states that only three kinds of bases have been given serious attention in this space, this includes B-Splines, cardinal splines and truncated power series splines \cite{de1972calculating}. We will discuss B-Splines in more depth later in this section, however Perperoglou et al \cite{perperoglou2019review} provide a good summary of cardinal and truncated power series splines, along with cubic and natural cubic splines. A cubic spline is where each basis function is a cubic polynomial and is bounded together at the inner knots, these are the most standard choice for polynomials as they appear smooth the human eye. A natural cubic spline takes into account the unpredictable behaviour of splines that can occur beyond the spline's domain, they therefore have properties that result in the polynomial pieces that are beyond data's boundaries that are linear. Cardinal spline basis are also a basis for a natural cubic spline in which the $j^{th}$ basis function is defined by its value at its specific knot and is undefined elsewhere. Finally the truncated power series spline are basis functions such that $S_1(x)=1, S_2(x)=x, S_3(x)=x^2, S_4(x)=x^3, S_5(x)=(x-\tau_1)^3_+, S_6(x)=(x-\tau_2)^3_+, ...$. Here the "+" subscript means that the this function should only be calculated if the argument $(x-\tau_1$) etc is positive, otherwise return zero. This means that as $x$ increase $S_5(x)$ only operates when $x$ reaches the first knot $\tau_1$, after that an extra cubic term is added. 

\subsubsection{\textsc{B-Splines}}
First proposed by Schoenberg (1946) \cite{schoenberg1946contributions}, B-Splines became an increasingly popular tool for mathematical smoothing in the 1970s following publications by De Boor \cite{de1972calculating} and Cox \cite{cox1978numerical} showcasing algorithms for computing B-Splines of any degree from B-Splines of a lower degree. B-Splines are highly attractive in non-parametric modelling and indeed Cox writes that B-Splines are 'eminently suitable for many numerical calculations'. Eilers and Marx \cite{eilers1996flexible} and Perperoglou \cite{perperoglou2019review} offer good summaries of the key properties and advantages of modelling with B-Splines and many authors build upon B-Splines to develop new techniques and methods of smoothing to represent complex models, we will explore these later in this section.

A B-Spline of degree $q$ consists of $q+1$ polynomial pieces each of degree $q$, which join together at $q$ inner knots. At the inner knots the derivatives up to $q-1$ are continuous and therefore provide a smooth function. The B-Spline is positive upon the support that is spanned over $q+2$ knots and is 0 everywhere else \cite{eilers1996flexible}, this provides the advantage of high numerical stability and makes them relatively simple to compute. Another advantage is that each spline function has bounded support. Take for example the truncated power term $(x-\tau_1)_+^3$, this is defined for all $x>\tau_1$. Remembering that in general $x$ can be as large as we like, we have to compute this term over all $x>\tau_1$ in our estimation procedure. However, B-Splines of degree 1 are each supported on just one interval, B-Splines of degree 2 are supported on at most two intervals and so on. We therefore only need to work them out on small subjects of the data, which helps with numerical calculations. 

B-Splines basis functions are also well-established tools within software packages in several coding languages, including R, MATLAB and Python, with over 13,000 R packages alone. These are provided in great detail in \textit{"A review of spline function procedures in R"}, Perperoglou et al (2019) \cite{perperoglou2019review}. In the examples you will see in this section, B-Splines basis are fitted to the data using the \texttt{Splines} library in R, specifically using function \texttt{bs} to generate a B-Spline basis matrix for representing piecewise polynomials. The number of knots and the degree of the polynomial pieces are specified by the user and are evaluated at a predictor value vector. The output of this procedure is a basis matrix with the dimensions of the length of the predictor value vector determined by the number of knots. 

Let us temporarily assume a model that is linear in selected spline functions of a single covariate, $x$. Further, we assume $n$ independent replications, so now for $i=1,2,...,n$
\begin{equation} \label{eq1}
    y_i = \sum_{j=1}^m \beta_j S_j(x_i) + \varepsilon_i
\end{equation}
where $\varepsilon_i$ is a zero mean error, and exactly one of the spline terms corresponds to an intercept. This is a standard linear model, which can be written in vector form
\begin{equation*}
    y = D\beta+\varepsilon
\end{equation*}
where $y$, $\beta$ and $\varepsilon$ are vectors of appropriate length, and $D$ is a design matrix with row $i$ corresponding to the spline vector of observation $i$. Note that we are not using any special notation for vectors. 

The standard least-squares estimator is
\begin{equation*}
    \hat{\beta} = (D^TD)^{-1}D^Ty
\end{equation*}
provided that $D^TD$ is invertible.

An advantage of expressing coefficient vector $\beta_j$ in Equation (\ref{eq1}) as linear is that we can interpret the estimation of $y$ as an optimisation problem in $S_j(x_i)$. This means that traditional estimation methods can be used for splines in generalized multivariable regression models \cite{perperoglou2019review}. 

\begin{figure}[H]
    \centering
    \includegraphics[width=1.0\textwidth]{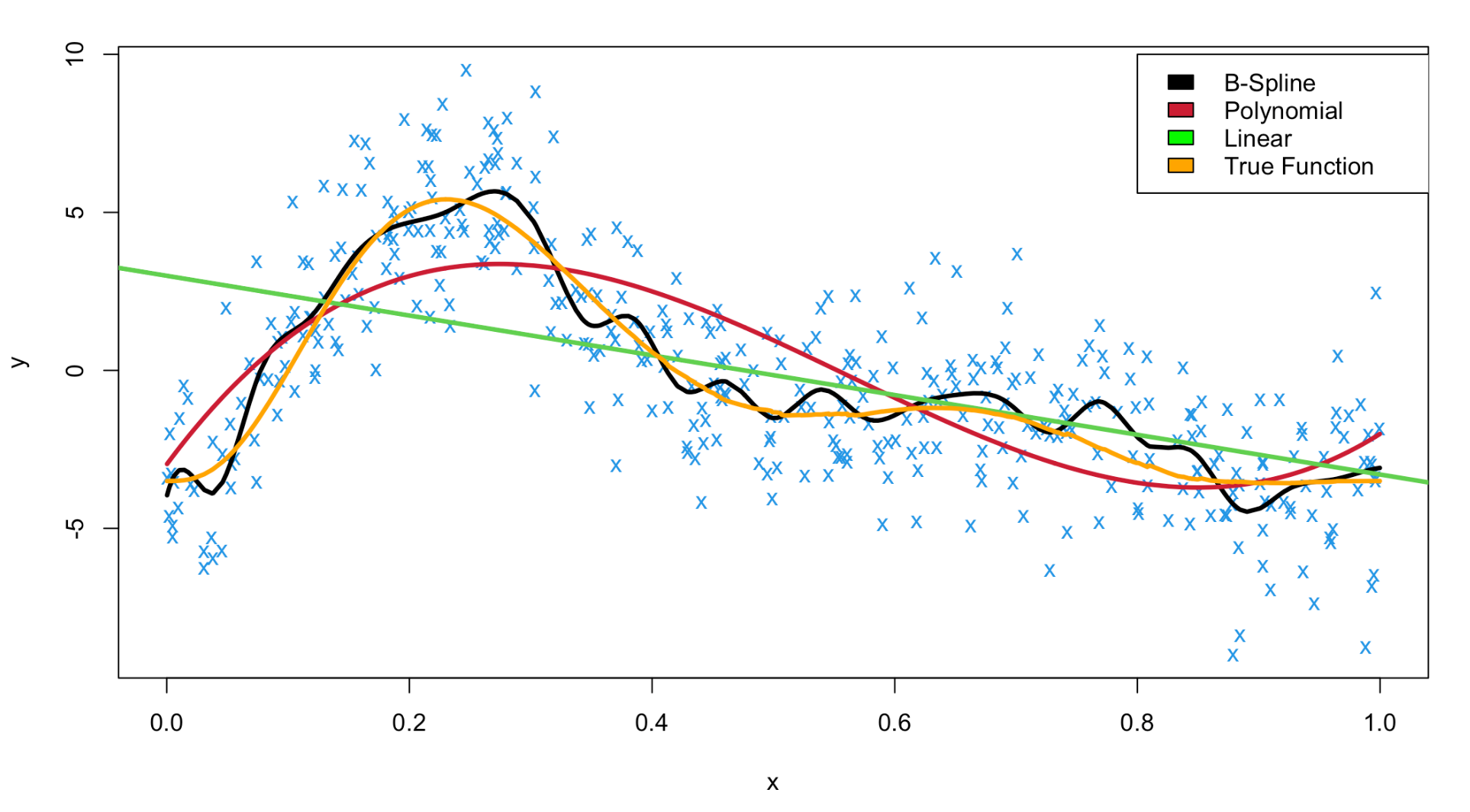}
    \caption{Comparison between linear, polynomial and spline fits to true function upon bivariate data. Green = linear function / red = polynomial function / black = B-Spline function / orange = true function}
    \label{fig:fig3}
\end{figure}

Figure \ref{fig:fig3} is an illustration of how a B-Spline can flexibly and robustly fit data in which there is no obvious linear or polynomial relationship. Figure \ref{fig:fig3} illustrates example bivariate data fitted within the $(x,y)$ plane and the fitting of three kinds of model: linear regression (green), polynomial regression (red) and B-Spline (black) along with the true relationship between $x$ and $y$ (orange). The data was created by generating 400 random samples from the uniform distribution for each covariate $x$ and $z$. The relationship the covariates $x$ and $z$ have with the true value for the response, denoted $y_{true}$ is found through an example, arbitrary equation:
\begin{equation*}
    y_{true} = -3.5 + 0.2x^{11}(10-10x)^6 + 10(10x)^3(1-x)^{10} - 1.8z^7(6-6z)^{5z^3}
\end{equation*}

Noise is then added to $y_{true}$ to give values for $y$ which are then plotted in Figure \ref{fig:fig3}, shown by the blue crosses. The three fits are then generated using R. The B-Spline fit is made up of polynomials of degree 3 hence they are cubic polynomials, and the number of knots is set to 50, this splits the domain [0,1] into 51 equidistant parts where a cubic spline is fitted within each subinterval and fused together at each knot by the conditions outlined above. Interestingly the polynomial fit is fitted also by using a B-Spline basis however the number of knots is set to zero, the result is ultimately a standard cubic polynomial fit. We see from Figure \ref{fig:fig3} that the B-Spline fit with 50 knots provides a far more flexible, robust and accurate modelling interpretation of the data, fitting more closely to the true function in orange than the linear or polynomial fits.

\begin{figure}[H]
    \centering
    \includegraphics[width=1.0\textwidth]{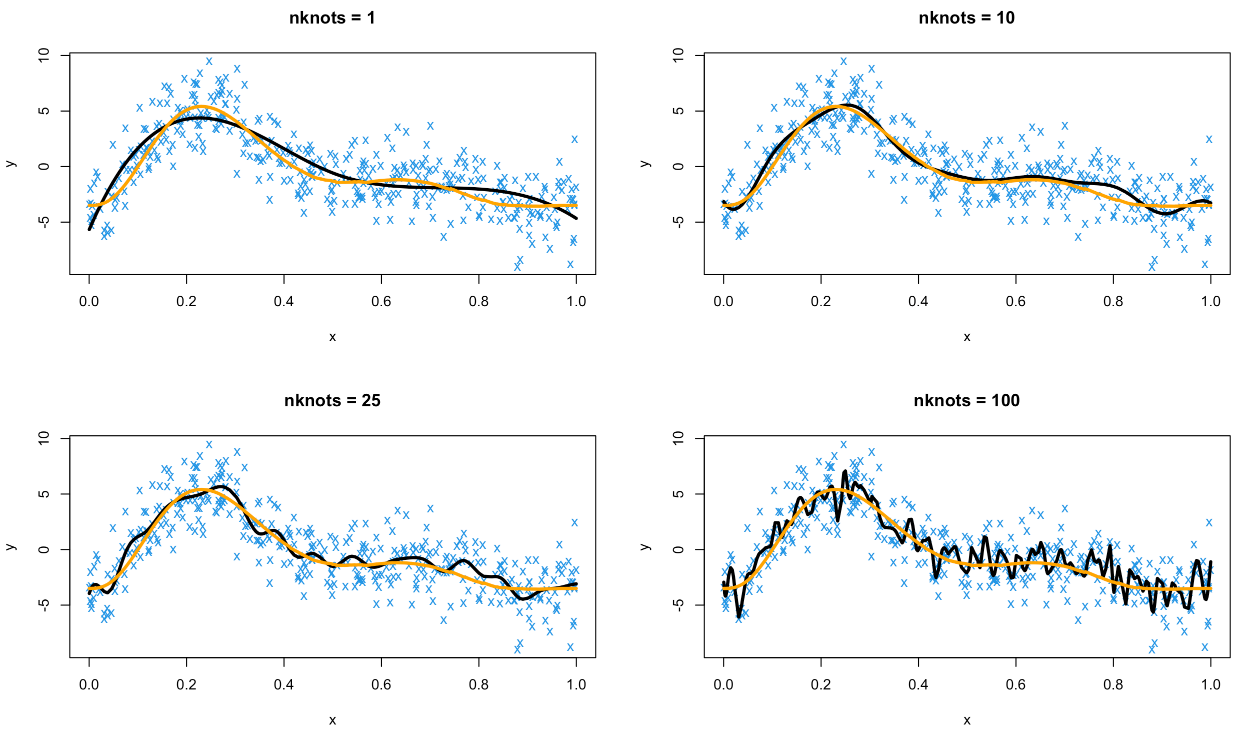}
    \caption{B-Splines fitted with varying number of knots. Orange line represents true function, black line represents fitted B-Spline}
    \label{fig:fig4}
\end{figure}

Using the same data generated to create Figure \ref{fig:fig3}, we highlight how changing the number of knots within the B-Spline fit can provide varying spline fits in Figure \ref{fig:fig4}. The number of knots determines the number of piecewise polynomials (in this case, of degree 3 hence cubic polynomials) are fused together to make the spline. Selecting the number of knots is important, too high number of knots can result in overfitting with high variance, whereas if the number of knots is too low this can result in an underfit with high bias where the relationship is not properly observed \cite{perperoglou2019review}. Figure \ref{fig:fig4} demonstrates four cubic splines with the number of knots varying from 1, 10, 25 and 100. We see from here that when the number of knots is equal to 1 the B-Spline presents an underfit when compared to the true function in orange. Conversely when the number of knots is equal to 100 there is a gross overfit to the data by the spline. The spline fits where the number of knots are equal to 10 and 25 provide a more appropriate fit of the real data. 

\subsubsection{\textsc{P-Splines (Penalized B-Splines)}}
So far we have considered only splines referred to as 'regression splines', i.e. unpenalized spline basis functions. Through adding a penalty term to these functions the result is what is known as 'smoothing splines'. Regression splines have their flexibility controlled by the number of knots as shown in Figure \ref{fig:fig4}, however as demonstrated a high number of knots may offer good flexibility but can result in an overfit to the data and likewise a low number of knots may offer a smooth but potential underfit; smoothing splines with their penalty term mean that less emphasis is required on the choice of the number and position of knots. Some works have focused upon methods of optimizing the number of knots selected in regression splines \cite{yeh2020fast,dung2017direct} however this is typically a difficult numerical problem and penalty terms can offer several numerical advantages.   

One example of a smoothing spline is a P-Spline, short for penalized B-Spline. B-Splines are formed sequentially and ordered, therefore for a smooth function we expect neighbouring coefficients to be similar; Eilers and Marx in their work \textit{"Flexible Smoothing with B-Splines and Penalties"} (1996) \cite{eilers1996flexible} proposed a penalty term based upon the higher order finite differences of these coefficients terms of adjacent B-Splines. This approach is a generalisation of O'Sullivan's work in 1986 who created a penalty based upon the second derivative of the fitted curve \cite{o1986automatic}. The formed objective function, i.e. the sum of squares (SS), is therefore represented as follows:

\begin{equation*}
    SS = \sum_{i=1}^n\biggl\{y_i-\sum_{j=1}^n\beta_j 
    S_j(x_i)\biggr\}^2 + \lambda \sum_{j=3}^m \biggl\{\Delta^2\beta_j\biggr\}^2 
\end{equation*}
in which the first term before the additive is the sum of squares between the observed data and the fitted B-Splines, and the second term after the additive is the penalty term, controlled by smoothing parameter $\lambda$. The $\lambda$ penalty term determines the level of smoothness that occurs, with smaller values of $\lambda$ resulting in a more jagged, rougher spline, and larger values leading to smoother, straighter curves. $\Delta$ is a difference operator with 
\begin{equation*}
    \Delta\beta_j = \beta_j - \beta_{j-1}
\end{equation*}
and $\Delta^2$ is the second order difference 
\begin{equation*}
\begin{split}
    \Delta^2\beta_j = \Delta(\Delta\beta_j) = \Delta\beta_j-\Delta\beta_{j-1} &= (\beta_j-\beta_{j-1})-(\beta_{j-1}-\beta_{j-2}) \\
    & = \beta_j-2\beta_{j-1}+\beta_{j-2}
\end{split}
\end{equation*}
The penalties are therefore squared linear combinations of the coefficients. We can collect the coefficients into a matrix C to give 
\begin{equation*}
    \sum^n_{j=3}\biggl\{\Delta^2\beta_j\biggr\}^2=\beta^TC^TC\beta, 
\end{equation*}
a quadratic in $\beta$, just as for the first term. It therefore follows that the sum of squares (SS) for a B-Spline with the Eilers and Marx higher order difference penalty is
\begin{equation*}
\begin{split}
    SS &= (y-S\beta)^T(y-S\beta)+\lambda\sum_j(\Delta^2\beta_j)^2 \\
    & = (y^T-\beta^TS^T)(y-S\beta)+\lambda(\beta^TC^TC\beta) \\
    & =y^Ty-2y^TS\beta+\beta^TS^TS\beta+\lambda(\beta^TC^TC\beta)
\end{split}
\end{equation*}
The estimated coefficients $\hat{\beta}$ can be found through differentiating and minimising the SS. From doing this we are able to find an equation for a fitted curve using a B-Spline with high order difference penalty:
\begin{equation*}
    \begin{split}
        \frac{\partial \mathrm{SS}}{\partial \beta} &= -2y^TS + 2S^TS\beta + 2\lambda C^TC\beta = 0 \\
    \end{split}
\end{equation*}
From this
\begin{equation*}
\begin{split}
    S^TS\beta + \lambda C^TC\beta = y^TS \\
    \implies \hat{\beta}(S^TS+\lambda C^TC) = y^TS \\
    \implies \hat{\beta} = (S^TS + \lambda C^TC)^{-1}y^TS
\end{split}
\end{equation*}
and so 
\begin{equation*}
    \hat{y} = S\hat{\beta}
\end{equation*}

\begin{figure}[H]
    \centering
    \includegraphics[width=1.0\textwidth]{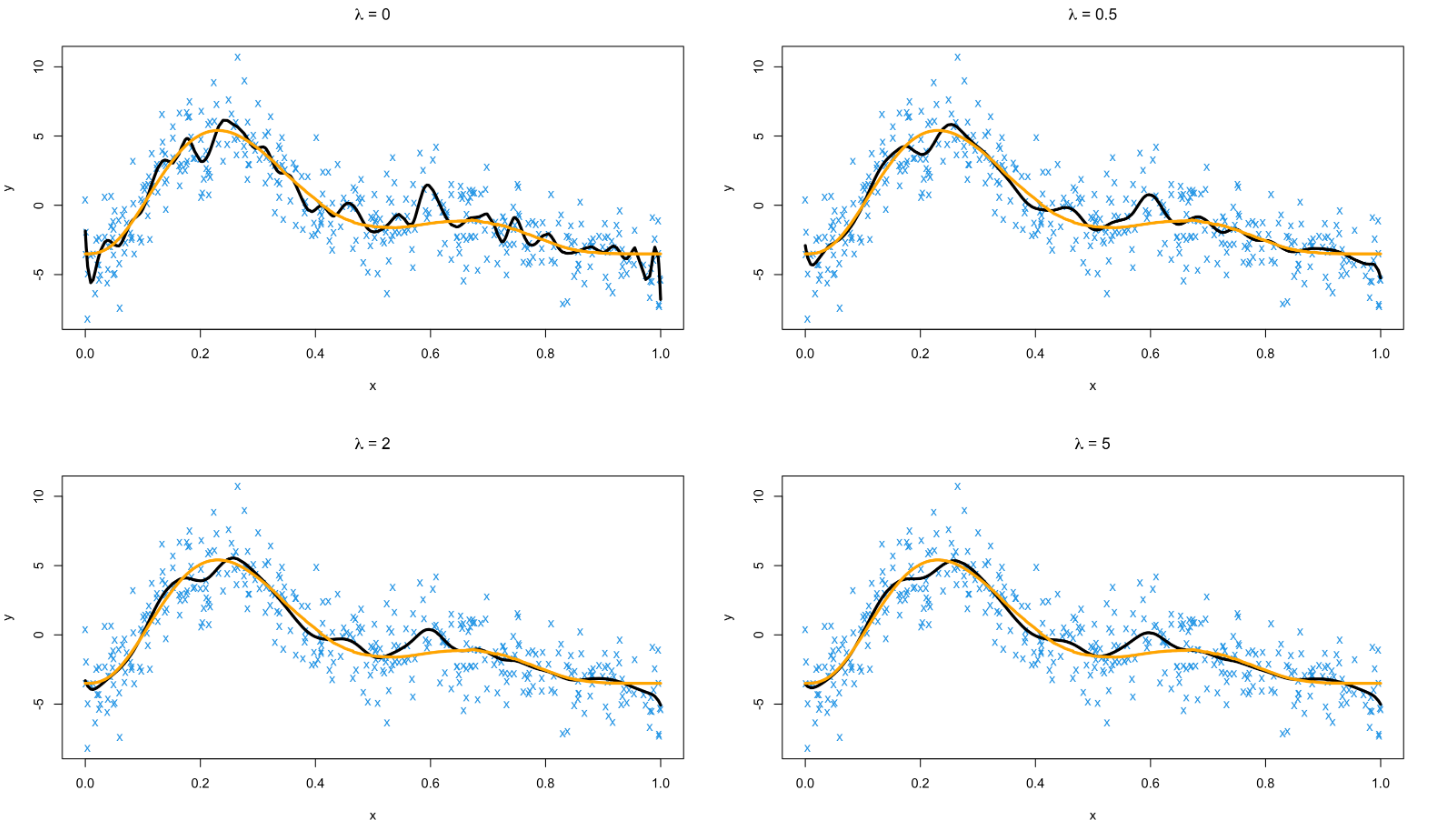}
    \caption{P-Splines fitted with varying magnitude of penalty term. Orange line represents true function, black line represents fitted P-Spline}
    \label{fig:fig5}
\end{figure}

Using the same data created within Figures \ref{fig:fig3} and \ref{fig:fig4}, we fit four P-Splines with varying magnitudes of penalty terms in Figure \ref{fig:fig5}. Note that the number of knots is set at 50 for each fit. 
As we can see from Figure 5, where $\lambda = 0$ and therefore no difference penalty term is applied, the resulting spline overfits the data and is particularly rough. As the penalty parameter $\lambda$ increases, the splines become smoother and get closer to the true function. Selection of an optimal penalty parameter is something discussed in later sections of this work. 

As their creators, Eilers and Marx provide several properties of P-Splines that make them particularly advantageous to use over the standard B-Spline. The key advantage is naturally the reduced need to focus on the number and position of knots that are necessary to create an appropriate fit to data; the implementation of P-Splines is encouraged through selecting a large number of knots and then simply using $\lambda$ to control the level of smoothness within the fitted curve \cite{currie2002flexible}. P-Splines also display no boundary effects, i.e. an erratic behaviour of the spline when modelled beyond a data's support, this is because the penalty term implements linearity constraints at the outer knots \cite{perperoglou2019review}. They are also able to conserve moments of the data, meaning that the estimated density's mean and variance will be equal to that of the data itself - often other types of smoothing such as a kernel smoother struggle to preserve variance to the data's level \cite{eilers1996flexible}. This property allows for valuable insights into the data's shape, distribution and central tendency.

The level of research that has been undertaken with P-Splines is extensive. This research covers a broad range of applications as well as adaptations to the P-Spline method itself, including modifications to the penalty term, as well as the addition of secondary penalty terms which this work also contributes to. P-Splines have been applied within many different domains, including in a medical context such as Mubarik et al (2020) applying P-Splines to breast cancer mortality data, managing to outperform existing non-smoothing models \cite{mubarik2020multi}, and also within a geospatial environment such as Rodriguez (2018) using P-Splines to model random spatial variation within plant breeding experiments, taking advantage of properties such as a stable and fast estimate, being able to handle missing data and being able to model a non-normal response \cite{rodriguez2018correcting}. P-Splines have also been adapted to be used within a Bayesian context by Lang and Brezger (2004) \cite{lang2004bayesian} and is shown in several works to improve predictive modelling, such as Brezger and Steiner's (2012) \cite{brezger2008monotonic} work of modelling demand for brands of orange juice and also Bremhorst and Lambert's (2016) work using survival analysis data \cite{bremhorst2016flexible}. 

There have also been several works that have built upon the original P-Spline method to incorporate a second additional penalty parameter, this has been done for a host of reasons. Aldrin (2006) introduces an additive penalty to the original Eilers and Marx P-Spline to improve the sensitivity of the smoothing curve \cite{aldrin2006improved}, whilst Bollaerts et al (2006) devises a second penalty to enforce a constraint in which the assumed shape of the relationship between predictors and covariates is taken into account \cite{bollaerts2006simple}. Simpkin and Newell (2013) also introduce a secondary penalty, suggesting this method helps alleviate fears when derivative estimation is of concern and can also lead to an improvement in the size of errors made during estimation \cite{simpkin2013additive}. Perperoglou and Eilers (2009) devise a second penalty term to capture excess variability and to explicitly estimate individual deviance effects; they use a ridge penalty to constrain these effects and the result is very effective, more suitable model than the single penalty P-Spline \cite{perperoglou2010penalized}. This work aims to contribute within the additionally penalized P-Spline method space, however the second penalty we use and our reasons for doing so are unlike that of other authors. We will discuss this more in the Model and Estimation section.

\section{\textsc{Model and Estimation}}
As mentioned in Section 1, the $\mathcal{H}$ dataset consists of $N_\mathcal{H}$ observations of response variable $y$ and two covariates $(x,z)$, and dataset $\mathcal{V}$ consists of $N_\mathcal{V}$ observations of response variable $y$ and single covariate $x$. We are interested in modelling a relationship between $y$ and the smooth function $\theta(x,z)$ represented using spline basis functions $S(.)$, which we can estimate from $\mathcal{H}$. However if $N_\mathcal{H}$ is small, there is therefore a lot of uncertainty surrounding this relationship. We therefore look to incorporate $\mathcal{V}$ to enhance our learning surrounding response variable $y$ and the relationship with covariates $(x,z)$. Provided that $\mathcal{V}$ is large, this will provide an accurate marginal estimate which can be incorporated into our analysis.

In this section we develop three models in which we are able to model this relationship. Our first model is a standard linear model with no penalization achieved using standard B-Splines, our second model is a P-Spline estimation building upon the linear B-Spline model with a single penalty term, and finally our third model is a proposed P-Spline estimation model with a new additional marginalisation penalty to aid with the incorporation of $\mathcal{V}$ into our smoothing relationship. Each model therefore builds upon the previous. We first of all introduce a series of model assumptions and simplifications, including for now, limiting our three models to utilising only a single covariate in $\mathcal{V}$ and two covariates in $\mathcal{H}$. It is important to note however that despite these simplifications this does not limit the overall contribution of our method to a real life dataset, which is explored more in details in the Applications section of this work. 

\subsection{\textsc{Model Assumptions}}
We assume in general that
\begin{equation*}
    g(E[y|x,z]) = \theta(x,z)
\end{equation*}
where $\theta(x,z)$ is a smooth function. For simulation purposes we take [0,1] to be the domain of each of $x$ and $z$, and we will use B-Splines to model $\theta(x,z)$. We do this in two ways:

\begin{enumerate}
    \item \textbf{No Interaction Between Covariates:} The relationship of the response $y$ to covariates $(x,z)$ treats each variable separately such that the model is made up of two smooth relationships. This is expressed as $\theta(x,z) = S(x) + S(z)$. 
    
    \item \textbf{Interaction Between Covariates:} There is a single smoothing relationship that incorporates an interaction of covariates $x$ and $z$ with response $y$. This is expressed as $\theta(x,z) = S(x,z)$.
\end{enumerate}

Each relationship results in different ways in which the design matrix of the B-Spline basis function is created which is explained in more detail in the Appendix to this paper. We will now introduce our three models for modelling the relationship between  $y$ and $\theta(x,z)$ starting with a linear B-Spline model, before building up to a single penalty P-Spline estimation model and then our proposed P-Spline model with a new additional marginalisation penalty.

\subsection{\textsc{Linear Model}}
In order to compare both the effect of fitting a smooth relationship between $y$ and covariates $(x,z)$, as well as the effect of our novel model we will introduce later, we first must have a means of comparison to the linear and once penalized methods we claim to outperform. Let us firstly assume a standard linear model:
\begin{equation*}
    y = D\beta + \varepsilon
\end{equation*}
where $D$ is the design matrix, $\beta$ are the corresponding coefficients and $\varepsilon$ are $N[0,\sigma^2]$ random errors as usual. We are able to find estimated values for our coefficients $\hat{\beta}$ by minimising the SS equation:
\begin{equation*}
    SS_0 = (y - D\beta)^T(y-D\beta),
\end{equation*}
ultimately receiving the ordinary least squares estimate:
\begin{equation*}
    \hat{\beta}_0 = (D^TD)^{-1}D^Ty
\end{equation*}
This value can then be used to receive fitted values for the linear model:
\begin{equation*}
    \hat{y} = D\hat{\beta}_0
\end{equation*}

Note that within this paper the standard linear model is denoted as model number 0, and therefore relevant variables and mathematical notation are denoted with a '0' subscript. 

\subsection{\textsc{P-Spline Estimation}}
Building upon the linear model and referring back to the penalty term described by Eilers and Marx in \textit{Flexible Smoothing with B-Splines and Penalties} (1996) \cite{eilers1996flexible}, we now apply a penalty to the B-Spline, known as a P-Spline estimation using a fairly large number of knots to create basis matrices $B_x$ and $B_z$. We denote $P_1$ and $P_2$ to be roughness matrices that are based upon the second-order differences in row and column directions, with $P_1$ referring to covariate $x$, and $P_2$ referring to covariate $z$. The construction of roughness matrices are discussed in more detail in the Appendix of this paper.

The least penalized squares estimate is now found through minimising:
\begin{equation*}
    \begin{split}
    SS_1 &= (y-D\beta)^T(y-D\beta)+\lambda_1(\beta^TP_1^TP_1\beta + \beta^TP_2^TP_2\beta)\\
    &= SS_0 + \lambda_1(\beta^TP_1^TP_1\beta + \beta^TP_2^TP_2\beta)
    \end{split}
\end{equation*}
The least penalized squares estimate is now
\begin{equation*}
    \hat{\beta}_1 = \biggl({D^TD+\lambda_1(P_1^TP_1+P_2^TP_2)\biggr)}^{-1}D^Ty
\end{equation*}
This value as previous can then be used to obtain the fitted value for the P-Spline estimation model:
\begin{equation*}
    \hat{y} = D\hat{\beta}_1
\end{equation*}
\begin{proof}
    Let $\Omega = P_1^TP_1 + P_2^TP_2$. The penalized sum of squares is\\
    \begin{equation*}
    \begin{split}
         PSS &= (y-D\beta)^T(y-D\beta)+\lambda_1\beta^T\Omega\beta \\
         &= y^Ty - 2\beta^TD^Ty + \beta^T(D^TD+\lambda_1\Omega)\beta
    \end{split}
    \end{equation*}
    Differentiating:

    \begin{equation*}
    \begin{split}
        \frac{\partial PSS}{\partial \beta} &= -2D^Ty + 2(D^TD+\lambda_1\Omega)\beta
    \end{split}
    \end{equation*}
    leading to 
    \begin{equation*}
    \begin{split}
        \hat{\beta}_1 &= (D^TD+\lambda_1\Omega)^{-1}D^Ty
    \end{split}
    \end{equation*}
    provided the inverse exists.
\end{proof}

In the P-Spline estimation model, both roughness matrices $P_1$ and $P_2$ are regulated by the same penalty parameter $\lambda_1$. This assumes that for our case $x$ and $z$ are symmetrical when simulating the data and for simulation purposes keeps the model complexity simple, however in reality we would need two parameters. 

Note that within this paper the P-Spline estimation model is denoted as model number 1, and therefore relevant variables and mathematical notation are denoted with a '1' subscript.

\subsection{\textsc{New Additional Marginalisation Penalty}} \label{Section 3.4}
As of yet we have not introduced a method of being able to take into account the vertical dataset, $\mathcal{V}$; we now introduce a second additional penalty to aid with this task. 

Suppose $x_{test}$ is a vector of $x$ values of chosen length to provide a reasonable spread across $x$ domain whilst not being too high in dimension. Let $\theta_{true}(x_{test})$ be the true marginal function at $x_{test}$, such that
\begin{equation*}
    \theta_{true}(x_{test})=g(E[y|x_{test}]),
\end{equation*}
which can be estimated from our vertical data $\mathcal{V}$. We are also able to estimate these marginal values from our horizontal data $\mathcal{H}$. 

Let:
\begin{itemize}
    \item $\hat{y} = \hat{\theta}(x,z) = D\hat{\beta}$, a vector of size $N_\mathcal{H} \times 1$.
    \item $x_0$ be any element from $x_{test}$ (a scalar).
    \item $(x_i,z_i)$ be covariates for element $i$ within $\mathcal{H}$.
    \item $k(.)$ be a kernel function, which we take to be the probability density function of a normal distribution with mean = 0 and standard deviation = 1.
    \item $\sigma_k$ be a smoothing parameter.
\end{itemize}
 
A consistent estimator, i.e. converges on the true value when sample size tends to infinity, is therefore:

\begin{equation*}
    \hat{\theta}_{\mathcal{H}}(x_0) = \frac{\sum_{i \in \mathcal{H}}k\biggl(\frac{x_i-x_0}{\sigma_k}\biggr)\hat{\theta}(x_i,z_i)}{\sum_{i \in \mathcal{H}}k\biggl(\frac{x_i-x_0}{\sigma_k}\biggr)}
\end{equation*}

In vector arguments we can write:
\begin{equation*}
    \hat{\theta}_{\mathcal{H}}(x_{test}) = K\hat{\theta}(x,z)
\end{equation*}
where $K$ is a matrix made up of scaled $k(.)$ functions. Recalling $\hat{\theta}(x,z) = D\hat{\beta}$, therefore:
\begin{equation*}
    \hat{\theta}_{\mathcal{H}}(x_{test}) = K\hat{\theta}(x,z) = KD\hat{\beta} = W\hat{\beta}
\end{equation*}
say.

We wish for $\hat{\theta}_\mathcal{H}(x_{test})$, i.e. our estimated marginal values from $\mathcal{H}$ of $x$, to be as close as possible to $\theta_{true}(x_{test})$, i.e. the true marginal values from $\mathcal{V}$ of $x$. In practice of course, $\theta_{true}(x_{test})$ would be unknown, however we can estimate this from the vertical data using $\hat{\theta}_\mathcal{V}(x_{test})$. Conversely, we have assumed since $N_\mathcal{V}\gg N_\mathcal{H}$, the error in $\hat{\theta}_\mathcal{V}(x_{test})$ will be relatively small. Hence for simplicity, we use the true marginal $\theta_{true}(x_{test})$ rather than the estimator $\hat{\theta}_\mathcal{V}(x_{test})$ for now. Our additional penalty term now added to the least penalized squares estimate takes this into account. 

The new least penalized squares estimate is now found through minimising:
\begin{equation*}
    \begin{split}
        SS_2 &= SS_1 + \lambda_2\biggl(\hat{\theta}(x_{test})-\theta_{true}(x_{test})\biggr)^T\biggl(\hat{\theta}(x_{test})-\theta_{true}(x_{test})\biggr) \\
        &= SS_1 + \lambda_2\biggl(W\hat{\beta}-\theta_{true}(x_{test})\biggr)^T\biggl(W\hat{\beta}-\theta_{true}(x_{test})\biggr)
    \end{split}
\end{equation*}

And thus providing the twice least penalized squares estimate:
\begin{equation*}
    \hat{\beta_2} = \biggr(X^TX+\lambda_1(P_1^TP_1+P_2^TP_2)+\lambda_2W^TW\biggl)^{-1}\biggl(X^Ty+\lambda_2W^T\theta_{true}(x_{test})\biggr)
\end{equation*}

This value as previous again can be used to obtain the fitted value for the P-Spline estimation model now fitted with an additional marginalisation penalty to take into account $\mathcal{V}$:
\begin{equation*}
    \hat{y} = D\hat{\beta}_2
\end{equation*}

\begin{proof}
    Let $\Omega = P_1^TP_1 + P_2^TP_2$. The twice penalized sum of squares is\\
    \begin{equation*}
         PSS = (y-D\beta)^T(y-D\beta) + \lambda_1\beta^T\Omega\beta + \lambda_2\biggl(W\beta-\theta_{true} (x_{test})\biggr)^T\biggl(W\beta-\theta_{true}(x_{test})\biggr)
    \end{equation*}
    Differentiating:
    \begin{equation*}
    \begin{split}
        \frac{\partial PSS}{\partial \beta} &= -2\biggl(D^Ty+\lambda_2W^T\theta_{true}(x_{test})\biggr)+2\biggl(X^TX\beta+\lambda_1\Omega\beta+\lambda_2W^TW\beta\biggr)
    \end{split}
    \end{equation*}
    leading to
    \begin{equation*}
    \begin{split}
        \hat{\beta_2} &= \biggl(D^TD+\lambda_1\Omega+\lambda_2W^TW\biggr)^{-1}\biggl(D^Ty+\lambda_2W^T\theta_{true}(x_{test})\biggr)
    \end{split}
    \end{equation*}
    provided the inverse exists.
\end{proof}

In this model we note that our additional marginalisation penalty is regulated by penalty parameter $\lambda_2$, and our P-Spline smoothing penalty is regulated by $\lambda_1$ as previously.

Within this paper the model in which we use the additional marginalisation penalty is denoted as model number 2, and therefore relevant variables and mathematical notation are denoted with a '2' subscript. We have now got three models where we are able to estimate coefficients $\hat{\beta}$ and therefore receive fitted values $\hat{y} = \hat{\theta}(x,z) = D\hat{\beta}$. It is now necessary to test each model upon simulated data to ensure that fitted values from our model that take into account the additional penalty, and thus the vertical data $\mathcal{V}$ are closer to the true values for $y$ - i.e. the model with the novel second penalty outperforms the linear and P-Spline estimation models. Note that this is not strictly a non-parametric model due to the assumption of a parametric model which is linear through the splines, formally however this is a 'flexible model' often referred to as 'non-parametric'.

\section{\textsc{Model Testing}}
In this section we test our three models upon a series of data simulations. Simulations allow for the exploration of a controlled space and also the freedom to adapt our models to a range of different parameters, including sample size, data noise and relationships between covariates. Using data simulations also allows for the use of perfect knowledge of true values for response variable $y$ and true marginal values of $x$, this provides the advantage that we are accurately able to compare our three models by comparing our fitted values of each model to the ground truth, something that is naturally not known in real world data. The aim of these simulations is to show that our adapted model featuring the additional marginalisation penalty outperforms both the linear B-Spline method and once penalized P-Spline estimation.

\subsection{\textsc{Data Simulation}}
\subsubsection{\textsc{Simulating Covariates and Response}}
We first of all generate some artificial $\mathcal{H}$ data. For now we generate $N_\mathcal{H} = 400$ observations and define $x$ and $z$ within the domain $[0,1]$. Covariates $x$ and $z$ are both distributed upon a regular grid spanning $(0,1)^2$. The relationship the covariates have with response variable $y$ depends upon whether we consider $x$ and $z$ to have independent effects (no interaction) from one another or not (interaction), therefore there are two separate equations, one to represent each model structure. The equations for these bivariate datasets are from Wood's \textit{Thin Plate Regression Splines} (2003) \cite{wood2003thin}. When we assume a model structure of no interaction effect, the true value for $y$, denoted $y_{true}$ is found through the equation:
\begin{equation}\label{eq:1} \tag{2}
    y_{true} = \frac{0.75}{\pi\sigma_x\sigma_z}\exp\biggl\{-(x-0.2)^2/\sigma_x^2-(x-0.3)^2/\sigma_z^2\biggr\} + \frac{0.45}{\pi\sigma_x\sigma_z}\exp\biggl\{-(z-0.7)^2/\sigma_x^2-(z-0.8)^2/\sigma_z^2\biggr\}
\end{equation}

When we assume a model structure with an interaction effect, $y_{true}$ is found through the equation:
\begin{equation}\label{eq:2} \tag{3}
        y_{true} = \frac{0.75}{\pi\sigma_x\sigma_z}\exp\biggl\{-(x-0.2)^2/\sigma_x^2-(z-0.3)^2/\sigma_z^2\biggr\} + \frac{0.45}{\pi\sigma_x\sigma_z}\exp\biggl\{-(x-0.7)^2/\sigma_x^2-(z-0.8)^2/\sigma_z^2\biggr\}
\end{equation}

Equations (2) and (3) are almost identical, the only difference being that when there is an interaction each exponent contains both $x$ and $z$, and when there is no interaction one exponent contains just $x$ and the other just $z$. Both relationships are each evaluated at $\sigma_x = 0.3$ and $\sigma_z = 0.4$. The value for $y$ is provided through adding artificial noise generated by $N_\mathcal{H}=400$ independent $N[0,\sigma^2]$ random variables to the $y_{true}$ values, which for now we evaluate at $\sigma = 0.2$. We display graphically the two relationships between covariates $(x,z)$ and response $y$ in a contour plot (Figure \ref{fig:fig6}) and as a 3D perspective plot (Figure \ref{fig:fig7}):
\begin{figure}[H]
    \centering
    \includegraphics[width=1.0\textwidth]{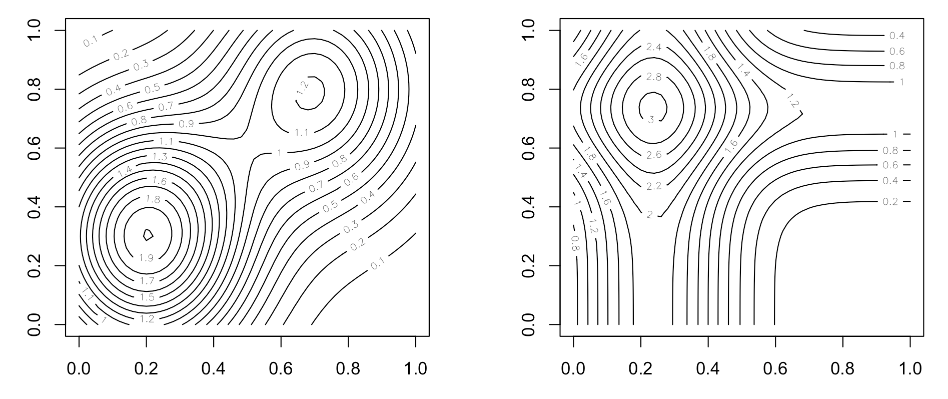}
    \caption{Contour plot of fitted relationships between $(x,z)$ and $y$ as model structure varies. \textit{Left: $\theta(x,z) = s(x,z)$ (interaction), Right: $\theta(x,z) = s(x)+s(z)$ (no interaction)}}
    \label{fig:fig6}
\end{figure}

\begin{figure}[H]
    \centering
    \includegraphics[width=1.0\textwidth]{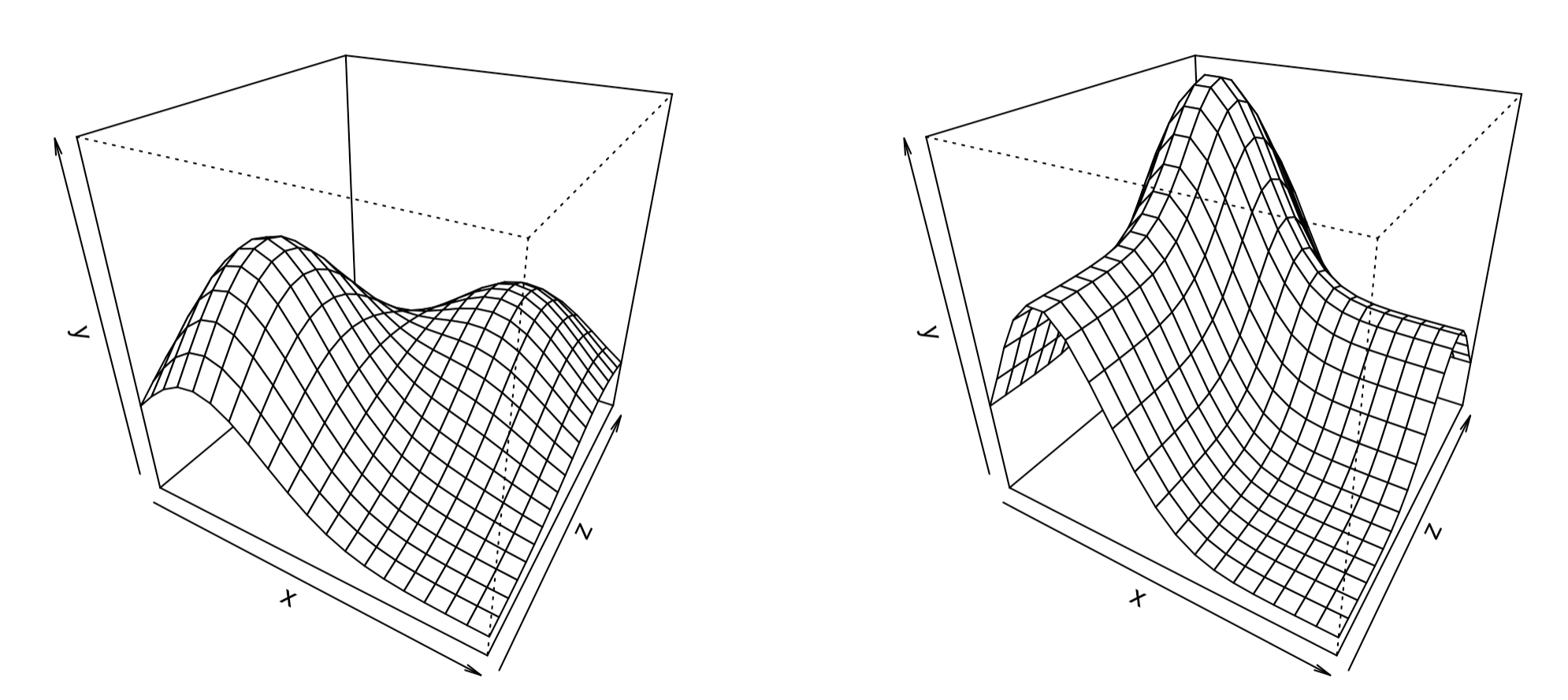}
    \caption{Perspective plot of fitted relationships between $(x,z)$ and $y$ as model structure varies. \textit{Left: $\theta(x,z) = s(x,z)$ (interaction), Right: $\theta(x,z) = s(x)+s(z)$ (no interaction)}}
    \label{fig:fig7}
\end{figure}

\subsubsection{\textsc{Estimating Marginal Effects}}

We next need to find $\theta_{true}(x_{test})$ in order to form our second penalty term. In our simulations take $x_{test}$ to be an equidistant sequence of 100 values between [0,1] to provide a reasonable spread across $x$ domain whilst remaining low in dimension. We estimated $\theta_{true}(x_{test})$ by calculating $y_{true}$ using either Equations (\ref{eq:1}) or (\ref{eq:2}) as appropriate, at each value of $x_{test}$ using 10,000 values of $z$ equidistant between [0,1] and then averaging. The value of 10,000 was selected upon to again provide good spread across $z$ domain and also to be large enough to provide an accurate true estimate. In this way we estimate $\theta_{true}(x_{test})$ under the assumption that $z$ is uniformly distributed. For other distributions we would need a weighted average. 

\begin{figure}[H]
    \centering
    \includegraphics[width=1.0\textwidth]{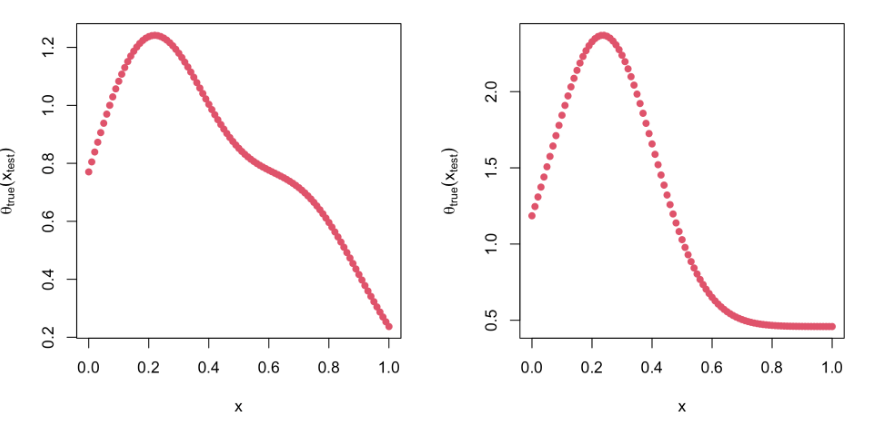}
    \caption{True marginal $\theta_{true}(x_{test})$ for the two model structures. \textit{Left: $\theta(x,z) = s(x,z)$ (interaction), Right: $\theta(x,z) = s(x)+s(z)$ (no interaction)}}
    \label{fig:fig8}
\end{figure}

Figure \ref{fig:fig8} is an illustration of the true marginal values $\theta_{true}(x_{test})$ for both relationships outlined in Section 4.1.1. 

\subsubsection{\textsc{Assessing Model Fit}}

To assess model fit we compare the fitted marginal of $\hat{\theta}(x)$ attained by our models with the true marginal of $x$ found in $\mathcal{H}$, $\theta_{true}(x)$, and we also compare the fitted values $\hat{y}$ of each model with the true values for $y$, $y_{true}$. The comparison for each case is in the form of sum of squares (SS), i.e. $\sum\{ \hat{y}-y_{true} \}^2$. The desired value is for this sum to be as close to zero as possible, as this will suggest a better fit. Note that in practise $y_{true}$ and $\theta_{true}(x)$ would be unknown, however for model testing/simulation purposes we assume that we have perfect knowledge. 

To summarise from our data simulation, we have attained $N_\mathcal{H} = 400$ values of $x,z$, as well as $N_\mathcal{H}$ values of $y$ and $y_{true}$. We also achieve the B-Spline design matrix $D$ as well as the true marginal effect of $x$ from our data $\theta_{true}(x)$. We are now in the position where we can assess the fit of all three model types.

\subsection{\textsc{Model Fit Comparison}}
We will first of all fit our three models to a single simulated dataset with a predetermined number of observations, level of noise and relationship between covariates $x$ and $z$ using the sum of squares of fitted values and sum of squares of marginal values as a means of comparison. We will also display graphically the three model fits alongside the ground truth for this simulated dataset. Following this we will then vary our simulated data's parameters and increase the number of simulations for each varying parameter combination. This exercise will illustrate that our model that utilises the additional marginalisation penalty outperforms the linear B-Spline and P-Spline estimate in both the singular data simulation case and the multiple simulation cases when parameters are varied. 

\subsubsection{\textsc{Single Dataset}}

The following three model fits are applied to a simulated dataset evaluated at $N_\mathcal{H} = 400$ and $\sigma = 0.2$. The data follows a structure where there is an interaction between $x$ and $z$ for now i.e. the effects are not independent. The number of knots for each covariate are set at the highest even number they can be at $p_x = p_z = 18$, noting that the restriction for this is that $p_xp_z + 1 < N_\mathcal{H}$ in order to create a valid design matrix. Finally penalty parameters are given default values of $\lambda_1 = 0.5$ and $\lambda_2 = 2$ for now - we will investigate later the optimal value for these penalty parameters. 

We denote Fit0 to be the initial linear model with B-Spline basis functions; Fit1 to be the P-Spline estimation model with smoothing penalty term $\lambda_1$; and Fit2 to be the P-Spline estimation model with new penalty term $\lambda_2$ to take into account the marginal value of $x$, along with $\lambda_1$ as previous. Recall as shown in Section 3 that the fitted values for each model are defined as follows:
\begin{equation*}
    \hat{y} = D\hat{\beta} 
\end{equation*}
in which:
\begin{equation*}
\begin{split}
    \text{Fit0:} \hspace{1cm} & \hat{\beta} = \hat{\beta}_0 = (D^TD)^{-1}D^Ty \\
    \text{Fit1:} \hspace{1cm} & \hat{\beta} = \hat{\beta}_1 = \biggl({D^TD+\lambda_1(P_1^TP_1+P_2^TP_2)\biggr)}^{-1}D^Ty \\
    \text{Fit2:} \hspace{1cm} & \hat{\beta} = \hat{\beta_2} = \biggr(D^TD+\lambda_1(P_1^TP_1+P_2^TP_2)+\lambda_2W^TW\biggl)^{-1}\biggl(D^Ty+\lambda_2W^T\theta_{true}(x_{test})\biggr)
\end{split}
\end{equation*}

And the fitted marginal values of $x$ for each model are defined as:
\begin{equation*}
    \hat{\theta}(x) = W\hat{\beta}
\end{equation*}
in which
\begin{equation*}
    W = KD
\end{equation*}
with $K$ defined in Section \ref{Section 3.4}.

We evaluate each model by finding the sum of squares for the fitted values 
\begin{equation*}
    \sum\biggl\{ \hat{y}-y_{true} \biggr\}^2
\end{equation*}
for all i's and the sum of squares for the marginal values of $x$
\begin{equation*}
    \sum\biggl\{ \hat{\theta}(x) - \theta_{true}(x) \biggr\}^2
\end{equation*}
for $x$ in $x_{test}$. The closer both of these sums are to zero, the better the model fit is to the data. In Table \ref{table:table1} we illustrate these sum of squares for each fit:
\begin{table}[H]
\centering
\begin{adjustbox}{width=0.5\textwidth}
\begin{tabular}{l|c|c|c}
\textbf{} &  \textbf{Fit0} & \textbf{Fit1} & \textbf{Fit2} \\ \hline
\textbf{SS Fitted Values} & 21.72 & 8.98 & 8.79 \\
\textbf{SS Marginal Values} & 0.50 & 0.52 & 0.27 \\
\end{tabular}
\end{adjustbox}
\caption{Sum of squares for model fits upon single dataset where $N_\mathcal{H} = 400, \sigma = 0.2$ and model structure is such that there is an interaction between $x$ and $z$}
\label{table:table1}
\end{table}

We see here that the sum of squares for the fitted values and the marginal values are both at their lowest for Fit2, the fit which incorporates the additional penalty term. Fit1 is also a better fit to the data than Fit0 in terms of sum of squares of fitted values, however Fit0 does display a better fit when comparing the sum of squares of marginal values.

We can illustrate the three fits upon the simulated data along with the true function in 3D plots in Figure \ref{fig:fig9}:
\begin{figure}[H]
    \centering
    \includegraphics[width=1.0\textwidth]{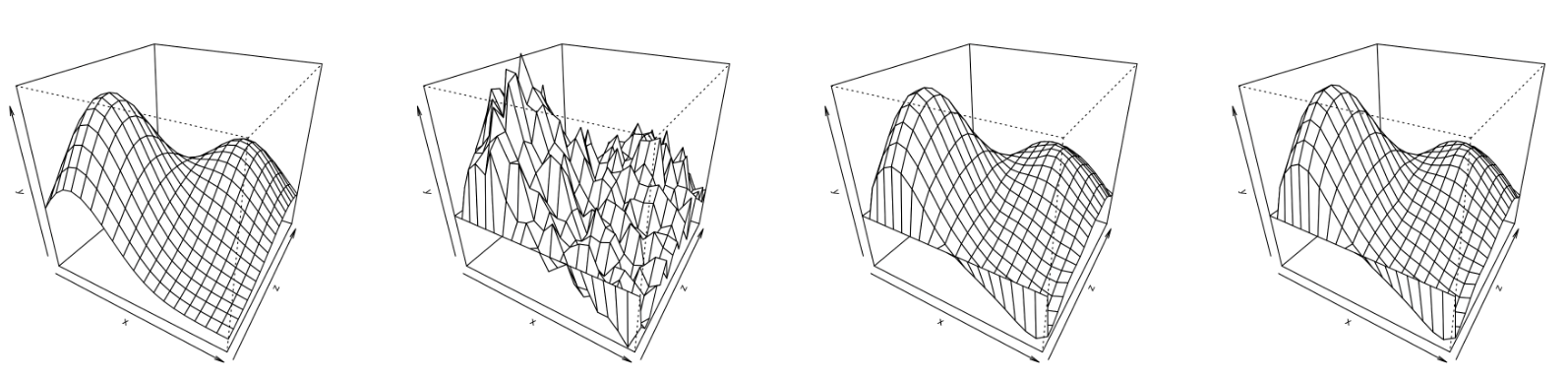}
    \caption{3D plot of all 3 model fits and the true function. From left to right: True function, Fit0, Fit1, Fit2.}
    \label{fig:fig9}
\end{figure}
We see that the linear model with B-Spline basis functions produces a very jagged fit between response $y$ and covariates $(x,z)$. When the penalty parameter $\lambda_1$ is introduced in Fit1 the fit becomes far smoother. It is difficult to spot any real difference between the model fit of Fit2 to Fit1 in the above 3D plots. 

\begin{figure}[H]
    \centering
    \includegraphics[width=1.0\textwidth]{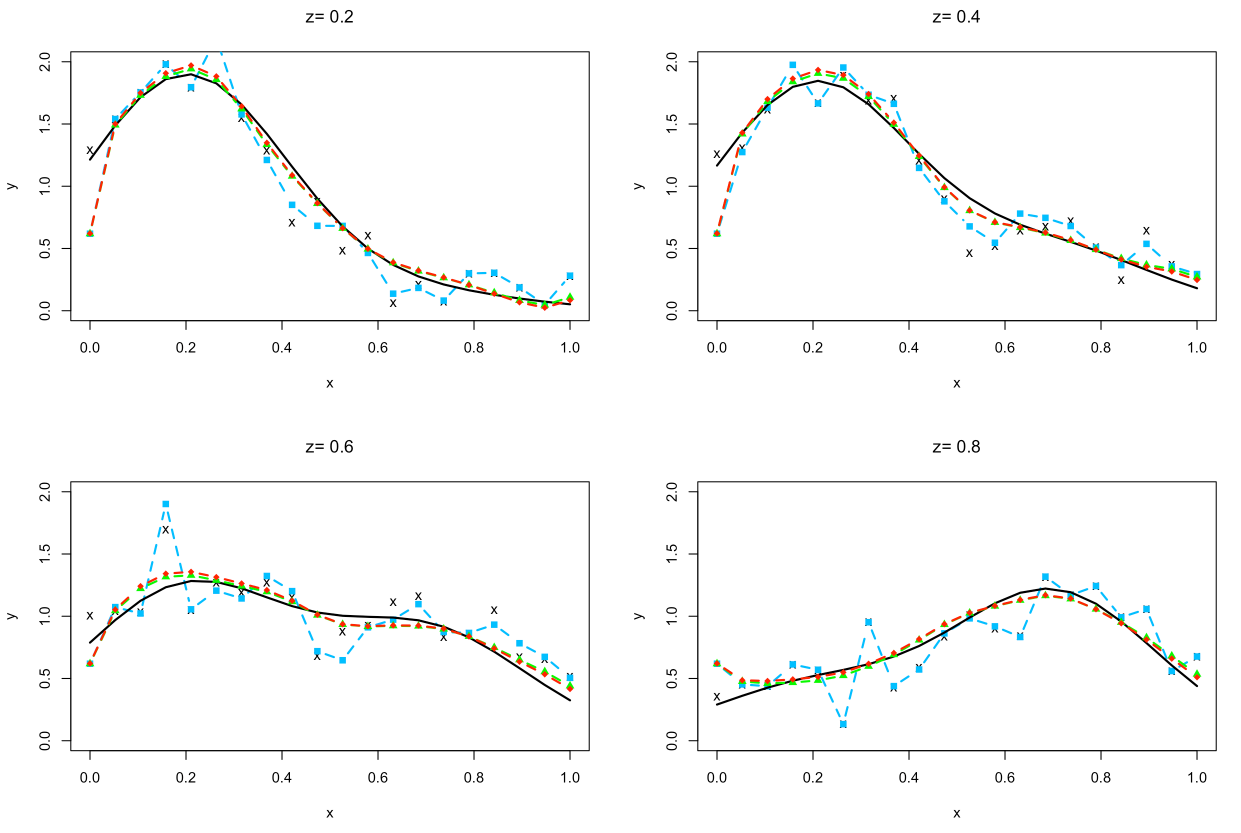}
    \caption{Black = True function $\theta(x,z)$ / Blue = Estimated model function $\hat{\theta}_\mathcal{H}(x,z)$ from Fit0 / Green = Estimated model function $\hat{\theta}_\mathcal{H}(x,z)$ from Fit1 / Red = = Estimated model function $\hat{\theta}_\mathcal{H}(x,z)$ from Fit2}
    \label{fig:fig10}
\end{figure}

Figure \ref{fig:fig10} illustrates for different values of $z$ the estimated model function from Fit0 (blue), Fit1 (green) and Fit2 (red), and the true function of the simulated data (black) between $x$ and $y$. These functions are demonstrated upon an $(x,y)$ plain when $z = 0.2, 0.4, 0.6 \text{ and } 0.8$.

We see from these plots that Fit1 and Fit2 offer a far better estimated function than Fit0, as illustrated by the green and red lines being closer to the black line, representing our true function. Generally for all values of $z$ the estimated function is quite jagged. Fit1 and Fit2 appear to offer similar fits and it is not visible on these graphs as to which is the better fit - only by looking at the sum of squares for the fitted values in Table 1 do we see that Fit2 is indeed an improvement upon Fit1. 

\begin{figure}[H]
    \centering
    \includegraphics[width=1.0\textwidth]{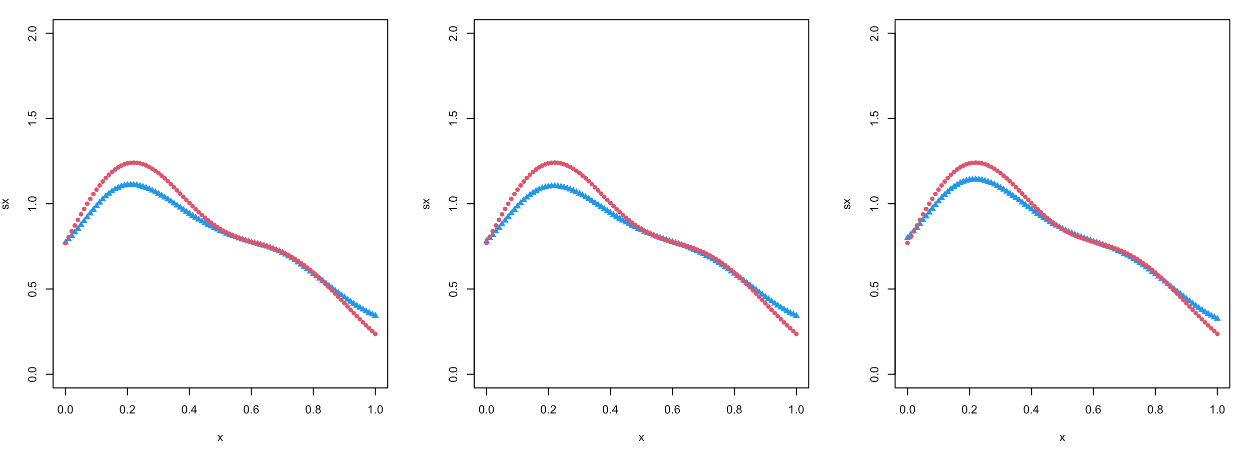}
    \caption{Blue = estimated marginal $\hat{\theta}_\mathcal{H}(x)$  / Red = true marginal $\hat{\theta}_\mathcal{H}(x_{test})$. From left to right: Fit0, Fit1, Fit2}
    \label{fig:fig11}
\end{figure}

Finally in Figure \ref{fig:fig11} we will plot the estimated marginal functions of $x$, $\hat{\theta}_\mathcal{H}(x_{test})$ found from each model fit against the true marginal fit $\theta_{true}(x_{test})$, recalling that in practise this would not be known. 
There is very little difference between the estimated marginal function in Fit0 and Fit1 however Fit2 does offer an improved fit to the true marginal function of $x$. This is highlighted in Table 1 where Fit2 has a lower sum of squares of the marginal values than that of Fit0 and Fit1.

We have shown that for one particular dataset where $N_\mathcal{H} = 400, \sigma=0.2$, in which there is an interaction between covariates, that Fit2 in which we use our novel additional penalty to take into account the marginal value for $x$, outperforms standard linear B-Spline methods and penalized P-Spline estimations in terms of model fit. This is however of course only one dataset and it is now necessary to investigate whether Fit2 works similarly better when we alter the number of observations $N_\mathcal{H}$, the level of noise in the dataset $\sigma$, and also when we test upon datasets where there is an interaction and no interaction between $x$ and $z$. It is also necessary to repeat these simulations many times to gauge whether these methods are working as expected. For each $N_\mathcal{H}, \sigma$ and model structure combination, the number of simulations is increased to 100, and the mean average sum of squares results for fitted and marginal values are displayed. 

\subsubsection{\textsc{Varying Size, Noise and Structure}}
Altering $N_\mathcal{H}$ to be = 100 or 400, $\sigma = 0.2, 0,5 \text{ or } 1.0$, and the relationship the two covariates have with one another, we display in Tables \ref{table:table2} and \ref{table:table3} the mean average sum of squares for fitted and marginal values from 100 simulations of each dataset with these varying combinations. Note that the number of knots, $p_x = p_z = 18$ for when $N_\mathcal{H}=400$ and $p_x = p_z = 8$ for when $N_\mathcal{H}=100$, recalling that $p_xp_z + 1 < N_\mathcal{H}$ must hold. Penalty parameters $\lambda_1$ and $\lambda_2$ are at their optimal values for each sample size, noise and covariate relationship combination - we will explore in later sections methods of calculating these values. 

\begin{table}[H]
\centering
\begin{adjustbox}{width=1\textwidth}
\begin{tabular}{ccc|ccc}
\textbf{Interaction} &  \boldmath{$N_\mathcal{H}$} & \boldmath{$\sigma$} & \textbf{Fit0 SS(Fitted)} & \textbf{Fit1 SS(Fitted)} & \textbf{Fit2 SS(Fitted)} \\ \hline
Yes & 100 & 0.2 & 6.55 & 4.74 & \textbf{4.71}\\
Yes & 100 & 0.5 & 19.95 & 7.73 & \textbf{7.34}\\
Yes & 100 & 1.0 & 70.26 & 15.41 & \textbf{13.20}\\
\hline
Yes & 400 & 0.2 & 20.64 & 9.07 & \textbf{8.99}\\
Yes & 400 & 0.5 & 88.66 & 13.18 & \textbf{12.53}\\
Yes & 400 & 1.0 & 333.85 & 23.46 & \textbf{20.32}\\
\hline
No & 100 & 0.2 & 0.68 & 0.54 & \textbf{0.45}\\
No & 100 & 0.5 & 4.23 & 3.04 & \textbf{2.08}\\
No & 100 & 1.0 & 16.84 & 10.47 & \textbf{5.88}\\
\hline
No & 400 & 0.2 & 1.56 & 0.75 & \textbf{0.68}\\
No & 400 & 0.5 & 8.98 & 3.43 & \textbf{2.67}\\
No & 400 & 1.0 & 37.76 & 11.65 & \textbf{8.19}\\
\end{tabular}
\end{adjustbox}
\caption{Mean average sum of squares of fitted values for each model fit as model structure, sample size and noise are varied (100 simulations)}
\label{table:table2}
\end{table}

\begin{table}[H]
\centering
\begin{adjustbox}{width=1\textwidth}
\begin{tabular}{ccc|ccc}
\textbf{Interaction} &  \boldmath{$N_\mathcal{H}$} & \boldmath{$\sigma$} & \textbf{Fit0 SS(Marginal)} & \textbf{Fit1 SS(Marginal)} & \textbf{Fit2 SS(Marginal)} \\ \hline
Yes & 100 & 0.2 & 0.94 & 0.98 & \textbf{0.73}\\
Yes & 100 & 0.5 & 1.71 & 1.51 & \textbf{0.74}\\
Yes & 100 & 1.0 & 4.93 & 3.53 & \textbf{0.76}\\
\hline
Yes & 400 & 0.2 & 0.32 & 0.40 & \textbf{0.22}\\
Yes & 400 & 0.5 & 0.56 & 0.67 & \textbf{0.20}\\
Yes & 400 & 1.0 & 1.34 & 1.22 & \textbf{0.16}\\
\hline
No & 100 & 0.2 & 0.81 & 0.95 & \textbf{0.53}\\
No & 100 & 0.5 & 1.89 & 2.13 & \textbf{0.58}\\
No & 100 & 1.0 & 4.77 & 4.73 & \textbf{0.48}\\
\hline
No & 400 & 0.2 & 0.75 & 0.83 & \textbf{0.67}\\
No & 400 & 0.5 & 0.90 & 1.10 & \textbf{0.64}\\
No & 400 & 1.0 & 1.95 & 2.35 & \textbf{0.62}\\
\end{tabular}
\end{adjustbox}
\caption{Mean average sum of squares of marginal values of $x$ for each model fit as model structure, sample size and noise are varied (100 simulations)}
\label{table:table3}
\end{table}

We see from Tables \ref{table:table2} and \ref{table:table3} that the average sum of squares values for both the fitted values and marginal values for $x$ that Fit2 is the lowest for every model structure, sample size and noise combination across 100 simulations. It is also the case that for all mean average sum of squares for fitted values that Fit1 outperforms Fit0 however when looking at mean average sum of squares for marginal values Fit1 does not always perform better than Fit0, this is not particularly surprising as there is nothing within the single penalty P-Spline that pulls the estimated marginal towards the true marginal. 

It is worth mentioning that comparing model fits between different combinations of parameters and structures is unwise. The key purpose of this exercise was to illustrate that when we take into account the marginal value of $x$ through the use of an additional penalty term, that this offers an improved model fit than that of existing linear and penalized regression methods. It is encouraging at this stage to see that Fit2 works as expected. In the next chapter we explore how we optimise the penalty parameters $\lambda_1$ and $\lambda_2$ for Fit1 and Fit2.

\section{\textsc{Optimizing Penalty Parameters in Simulations}}
In P-Splines, the larger $\lambda$ is, the more penalized the curvature of the fit is, therefore it is less sensitive to the data providing lower variance and higher bias. As $\lambda \to 0$, bias is low and variance is high. Typically we would want $N \to \infty$ as $\lambda \to 0$. Our problem is more complex to solve as previous literature offers solutions when there is only a single penalty parameter; in our case with Fit2 and our additional penalty term, we require the selection of two penalty terms, $\lambda_1$ and $\lambda_2$.

In the previous chapter, results displayed in Tables \ref{table:table2} and \ref{table:table3} for Fit1 and Fit2 are found through using the 'optimal' values for penalty parameters $\lambda_1$ and $\lambda_2$, for each model structure and parameter combination. This section explores how these parameters were found.

Focusing initially upon one dataset which considers an interaction between covariates $x$ and $z$, and model parameters $N_{\mathcal{H}} = 400$ and $\sigma = 1.0$, we elect to treat the penalty parameter $\lambda_1$ differently within Fit1 and Fit2 such that $\lambda_{1a}$ determines Fit1 only and $\lambda_{1b}$ determines Fit2 along with $\lambda_2$. 

Within Fit1, we aim to find the value of $\lambda_{1a}$ that minimises the value of the sum of squares between the fitted values, $\hat{y}$, and the true values of the response, $y_{true}$. We require $\lambda$ values to be non-negative, therefore we select an appropriate range of values for $\lambda_{1a}$, and fit our simulated dataset using Fit1 with each $\lambda_{1a}$ value within this range. The range selected in this case were values [0,1,2,...,100]. Note that in practice this is not possible as $y_{true}$ is unknown, this is discussed later. Figure \ref{fig:fig12} is a plot of each SS for fitted values corresponding to each value of $\lambda_{1a}$ within the specified range, with the minimum value of SS and corresponding $\lambda_{1a}$ value highlighted:
\begin{figure}[H]
    \centering
    \includegraphics[width=0.75\textwidth]{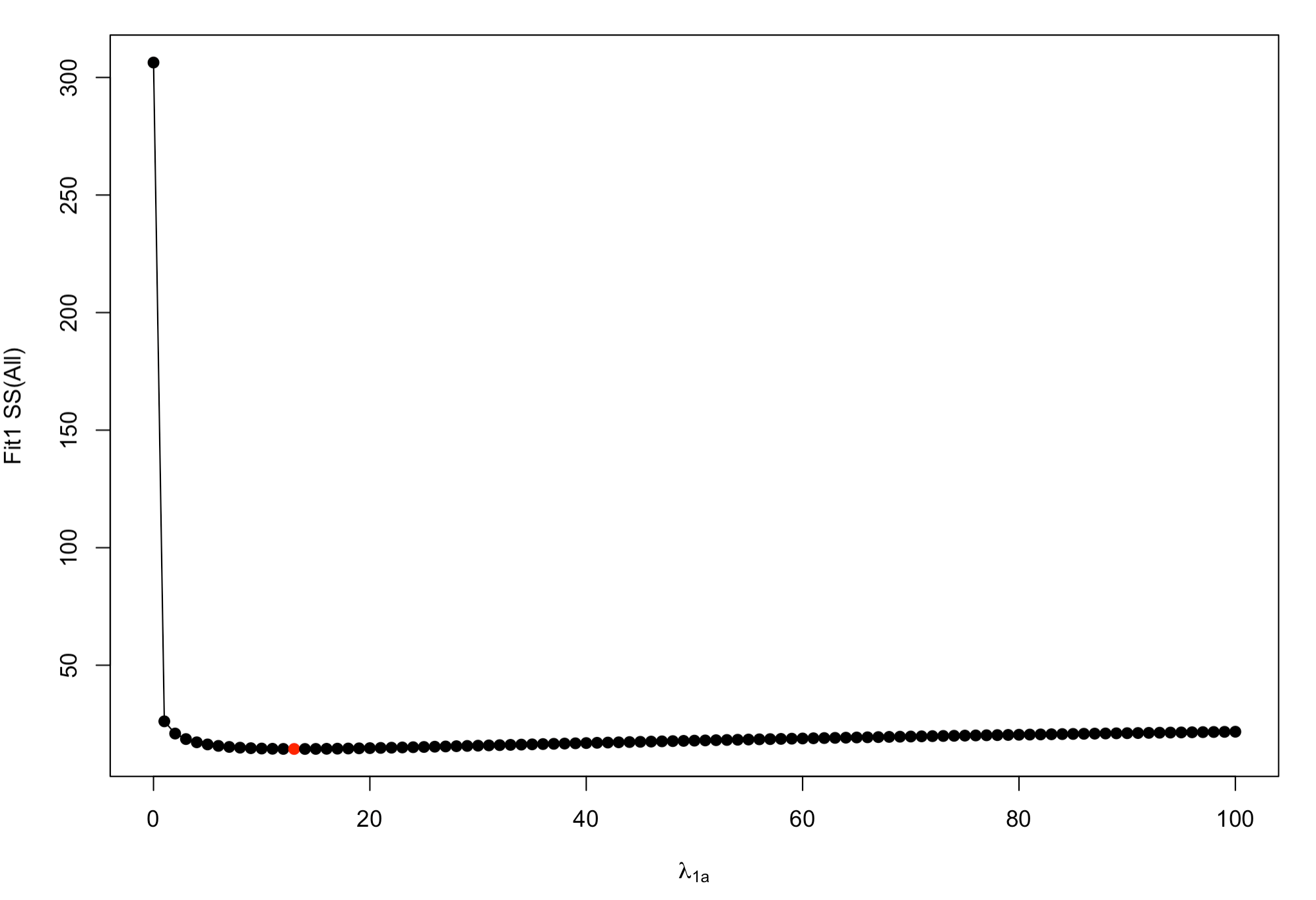}
    \caption{SS(Fitted) for Fit1 values for an interaction dataset between covariates where $N_\mathcal{H} = 400$ and $\sigma=1.0$, whilst varying $\lambda_{1a}$ within a [0,100] range.}
    \label{fig:fig12}
\end{figure}
From the 100 different values of $\lambda_{1a}$ used we find that a value of $\lambda_{1a}=13$ provides the lowest value of SS(Fitted) for Fit1 upon this particular dataset. We can see from this graph that the sum of squares for this fit is not very sensitive to changing $\lambda$ values after the first few values within our range. We now wish to find the 'optimal' values of $\lambda_{1b}$ and $\lambda_2$ that minimises the fitted sum of squares for Fit2. Fixing for now $\lambda_{1a}=\lambda_{1b}=13$, we define $\lambda_2$ to be along the range [0,0.5,1.0,...,50.0]. We now plot in \ref{fig:fig13} each SS value corresponding to each $\lambda_2$ value, with the minimum value highlighted:
\begin{figure}[H]
    \centering
    \includegraphics[width=0.75\textwidth]{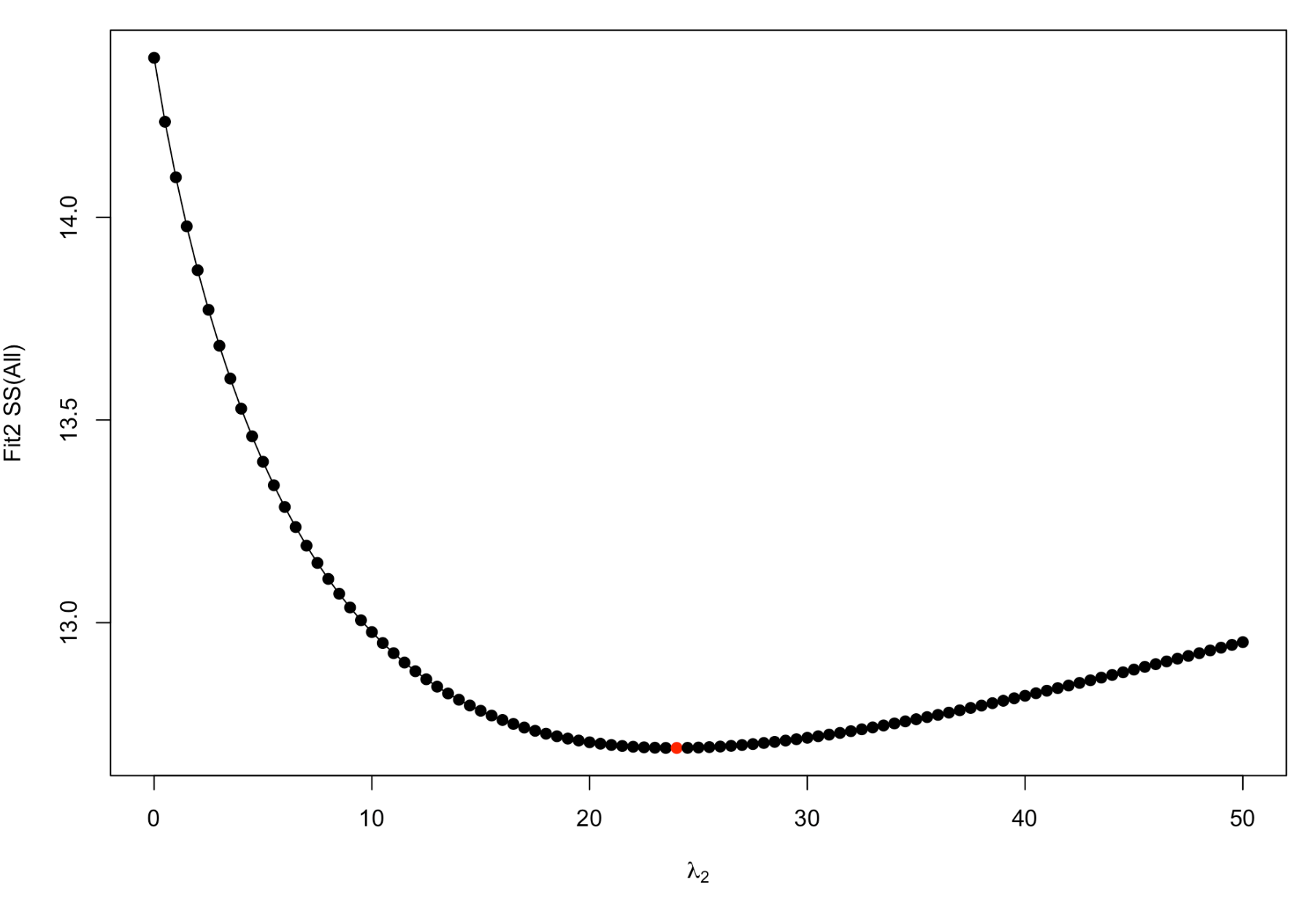}
    \caption{SS(Fitted) for Fit2 values for an interaction dataset between covariates where $N_\mathcal{H} = 400$ and $\sigma=1.0$, whilst varying $\lambda_{2}$ within a [0,50] range.}
    \label{fig:fig13}
\end{figure}

From this figure the $\lambda_2$ value that provides the optimum SS(Fitted) value for Fit2 when fixing $\lambda_{1a}=\lambda_{1b}=13$, is $\lambda_2 = 24$ for this specific dataset.

Finally we wish to find the optimum value of $\lambda_{1b}$ for Fit2, now that we know the optimum value for $\lambda_2 = 24$. Allowing $\lambda_{1b}$ to be defined along the same range as $\lambda_{1a}$, [0,1,...,100], we in the same way again plot the SS(Fitted) for Fit2 whilst varying $\lambda_{1b}$, highlighting the optimum value in Figure \ref{fig:fig14}:
\begin{figure}[H]
    \centering
    \includegraphics[width=0.8\textwidth]{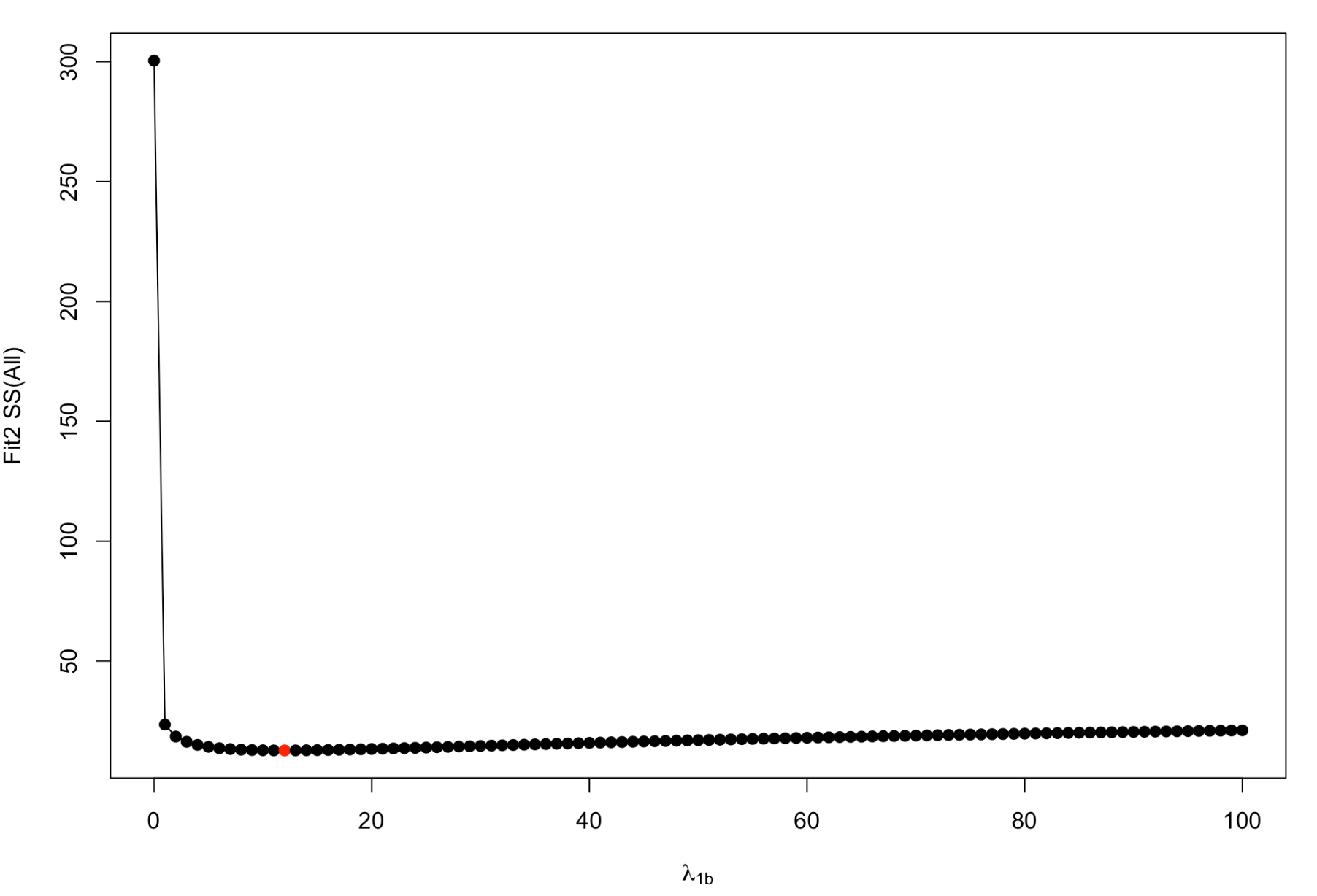}
    \caption{SS(Fitted) values for Fit2 values for an interaction dataset between covariates where $N_\mathcal{H} = 400$ and $\sigma=1.0$, whilst varying $\lambda_{1b}$ within a [0,50] range.}
    \label{fig:fig14}
\end{figure}
We have therefore found that for this particular dataset, the values for the penalty parameters are $\lambda_{1a}=13, \lambda_{1b}=12, \lambda_2=24$.  Of course this is only one dataset and it is therefore unwise to declare these penalty parameter values to be the optimum for every data simulation. It is also the case that as $N_{\mathcal{H}}$, $\sigma$ and covariate relationships are altered, these values could be drastically different. For each parameter and model structure combination we therefore repeat the process that we have outlined above 100 times, and then calculate the mean average for each penalty parameter across these 100 repetitions. It is also important to note that the penalty parameter values obtained from this search method are not necessarily the optimum values; to find these values we would require a two-dimensional search over ($\lambda_{1b},\lambda_2$) simultaneously. This has been omitted from this work as a simultaneous search would be very time consuming as this would need to be carried out for all 100 simulations across all parameter combinations. Our search method should therefore be considered as an approximation.  

Rounded to sensible values, we display the optimum $\lambda$ values for each model structure and dataset parameter combination in Table \ref{table:table4}.

\begin{table}[H]
\centering
\begin{adjustbox}{width=0.4\textwidth}
\begin{tabular}{ccc|ccc}
\textbf{Interaction} &  \boldmath{$N_\mathcal{H}$} & \boldmath{$\sigma$} & \boldmath{$\lambda_{1a}$} & \boldmath{$\lambda_{1b}$} & \boldmath{$\lambda_{2}$} \\ \hline
Yes & 100 & 0.2 & 0.1 & 0.1 & 0.2 \\
Yes & 100 & 0.5 & 0.3 & 0.3 & 0.6 \\
Yes & 100 & 1.0 & 0.9 & 0.9 & 1.8 \\
\hline
Yes & 400 & 0.2 & 2 & 2 & 2.3 \\
Yes & 400 & 0.5 & 6 & 6 & 7 \\
Yes & 400 & 1.0 & 18 & 18 & 21 \\
\hline
No & 100 & 0.2 & 0.1 & 0.1 & 0.5 \\
No & 100 & 0.5 & 0.3 & 0.3 & 1.5 \\
No & 100 & 1.0 & 0.9 & 0.9 & 4.5 \\
\hline
No & 400 & 0.2 & 4.3 & 6 & 1 \\
No & 400 & 0.5 & 13 & 18 & 3 \\
No & 400 & 1.0 & 36 & 54 & 9 \\
\end{tabular}
\end{adjustbox}
\caption{Penalty parameter values for each model structure and parameter combinations}
\label{table:table4}
\end{table}

We see generally that as $N_\mathcal{H}$ and $\sigma$ increase, the size of each penalty parameter increases also. For datasets with a covariate interaction the optimum values for $\lambda_1$ typically follow $\lambda_{1a}=\lambda_{1b}$, however for non-interaction datasets $\lambda_{1a}$ and $\lambda_{1b}$ differ when $N_\mathcal{H}$ is larger. Despite this method being particularly lengthy, the results produced as seen in Tables 2 and 3 demonstrate that with these optimised penalty parameters that Fit2 provides an improved fit upon Fit0 and Fit1. 

This is an ad-hoc method of finding the optimum penalty parameters, however the results that it provides with regards to sum of squares to fitted and marginal values as shown in Tables \ref{table:table2} and \ref{table:table3} in the previous section, it illustrates that Fit2 which we use our additional new penalty provides a better fit than that of Fit0 or Fit1. Within simulations this method of selecting penalty parameter values is valid as the sum of squares is a comparison between the fitted values and the true values of the response. In reality this is unknown and a new method of optimising $\lambda$ values is required. In the Applications section of this paper a cross-validation method is presented and utilised. 

\section{\textsc{Logistic Regression for a Binary Response}}
In previous works on predicting NAFLD related-endpoints of patients using only basic, easily accessible information, the response variables that were observed were all binary, i.e. either a person had reached a certain endpoint (1), or they hadn't (0). In order for the models we have created to now take into account a binary response $y$, several adaptations are required. 

Let us now assume now that $y$ is a binary response such that: 
\begin{equation*}
    Pr(Y = 1 | x,z) = \theta(x,z)
\end{equation*}
in which $\theta(x,z)$ is a smooth, yet unknown function of probabilities. We can calculate the marginal effect of $x$ via:
\begin{equation*}
    Pr(Y = 1 | x) = \theta(x) = \int \theta(x,z) f_z(z | x) dz
\end{equation*}

As we have done so in Section 4, we are able to estimate the smooth function $\theta(x,z)$ from our horizontal data $\mathcal{H}$. We now again modify our estimated smooth function so that the marginal estimate of $x$ from the horizontal data, $\hat{\theta}_\mathcal{H}(x)$ is close to the more accurate marginal found from the vertical data, $\hat{\theta}_\mathcal{V}(x)$. For now, we assume that $N_\mathcal{V}$ is so large that the uncertainty that comes from $\hat{\theta}_V(x)$ is so small that we may as well use the true marginal $\theta(x)$ instead. Reiterating that in reality this would not be possible as we would not know the true marginal, however for simulation purposes it is useful as we try and achieve a marginal estimate $\hat{\theta}_{\mathcal{H}}(x)$ as close to the truth as possible. 

\subsection{\textsc{Data Creation}}
For data creation in simulations, we allow the true probabilities to be equal to the standard logistic function, also known as the expit:
\begin{equation*}
    \theta(x,z) = \frac{e^{s(x,z)}}{1+e^{s(x,z)}}
\end{equation*}
in which $s(x,z)$ is a scaled form of the smooth function we used previously in Section 4 for the linear model simulations. Scaling ensures that the true probabilities are reasonable and spanned across a good range, without being all too near 0 or 1 as illustrated in Figure \ref{fig:fig15}. In simulated binary response data, $y$ is created through generating random samples from a uniform distribution. We find the true marginal $\theta(x)$ in our simulations through the same way as demonstrated in Section 4 also. True probabilities $\theta(x,z)$ are shown in Figure \ref{fig:fig15} for a non-interaction dataset where $N_\mathcal{H}$ and $p_x=p_z=8$:
\begin{figure}[H]
    \centering
    \includegraphics[width=0.8\textwidth]{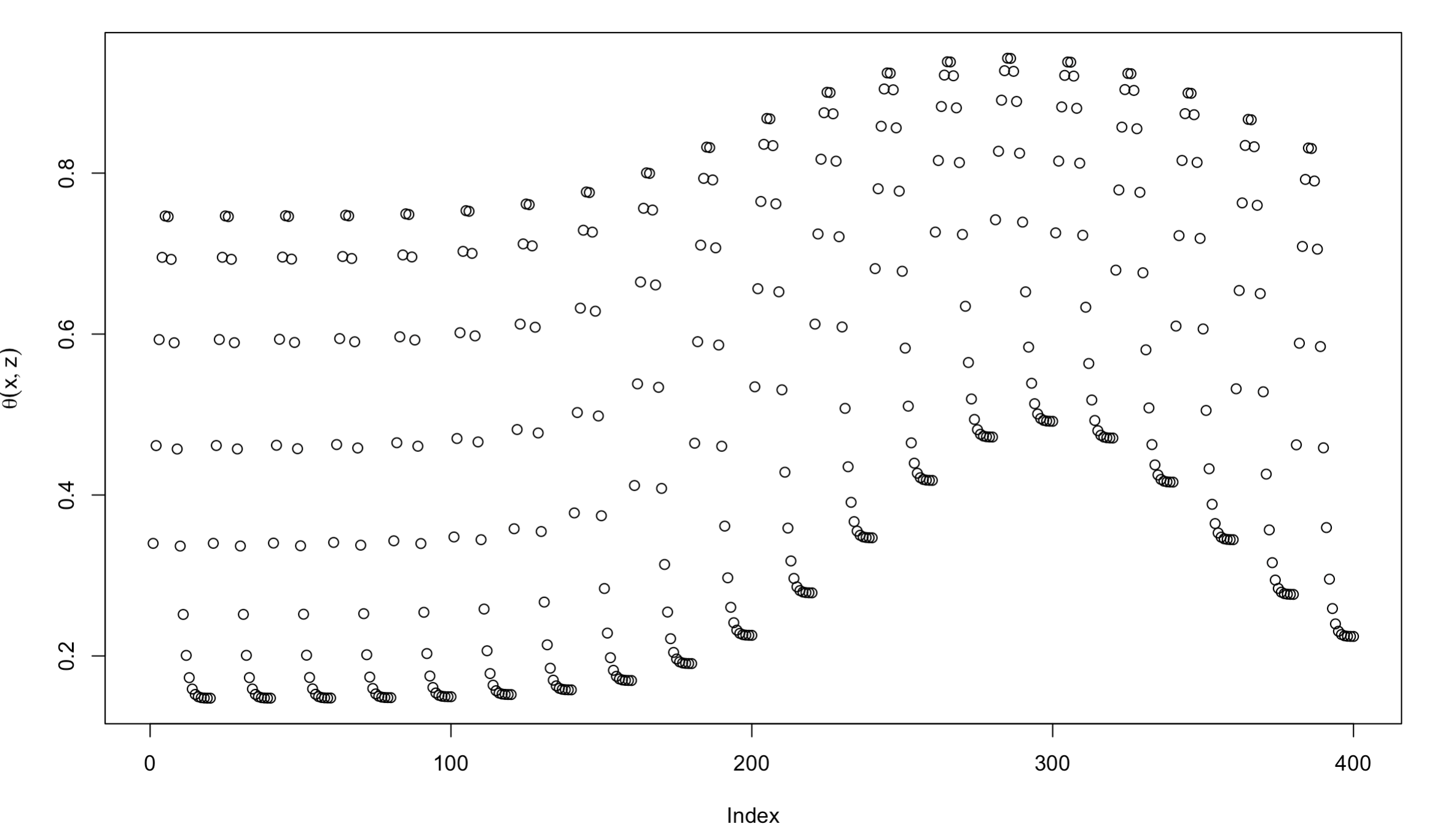}
    \caption{True probabilities $\theta(x,z)$ for non-interaction dataset where $N_\mathcal{H}=400$ and $p_x=p_z=8$.}
    \label{fig:fig15}
\end{figure}

As previous in the linear model, we will begin with a B-Spline approximation to create design matrix $D$ for the horizontal data $\mathcal{H}$, however now we will take the logistic model rather than the linear. Therefore for any case $i$:
\begin{equation*} 
    \theta(x_i,z_i) = \frac{e^{d_i^T\beta}}{1+e^{d_i^T\beta}}
\end{equation*}
in which $d_i^T$ is row $i$ in design matrix $D$. 

Now that we have acquired covariates $x$ and $z$, binary response $y$, true fitted probabilities $\theta(x,z)$, true marginal probabilities $\theta(x)$, and design matrix $D$, we are in the position where we can fit a standard logistic regression model (Fit0), followed by a P-Spline estimate as is standard to avoid overfitting (Fit1), and then by adding the novel marginal penalisation, penalising discrepancies in marginal values (Fit2). There are several differences that occur from the linear approach as seen previously which we will highlight in the following sections. 

\subsection{\textsc{Logistic Regression - No Penalties Added}}
The first difference in this approach is that we estimate our coefficient values $\hat{\beta}$ using maximum likelihood estimation rather than through least squares. Allowing $\theta_i = \theta(x_i,z_i)$ for $i \in \mathcal{H}$, the likelihood is:
\begin{equation*}
    L(\beta) = \prod_{i \in \mathcal{H}}\theta_i^{y_i}(1-\theta_i)^{1-y_i}
\end{equation*}
    
and the log-likelihood is:
\begin{equation*}
    l(\beta) = \sum_{i \in \mathcal{H}}\{y_i\log\theta_i + (1-y_i)\log(1-\theta_i)\}
\end{equation*}

Unfortunately there is no simple closed form for $\hat{\beta}$ which maximises $l(\beta)$, we therefore require a numerical method. As the first and second derivatives can be easily obtained the obvious choice is the Newton-Raphson method \cite{ypma1995historical}. 

Let us assume that design matrix $D$ is $N_\mathcal{H} \times p$ in dimension, and so 
\begin{equation*}
    \frac{\partial l}{\partial \beta} \text{   and 
  } \frac{\partial^2 l}{\partial \beta^2}
\end{equation*}
are a $p \times 1$ vector of first derivatives and $p \times p$ matrix of second derivatives respectively. Iteratively, we start with an initial coefficient estimate guess of $\beta_0$ and then create a sequence $\beta_1, \beta_2, ... $ until the sequence has converged, or is adjudged have converged sufficiently. 

Allowing the current estimate to be $\beta_k$, then the next estimate according to the Newton-Raphson method is defined as:
\begin{equation*}
    \beta_{k+1} = \beta_k - \biggl(\frac{\partial^2 l}{\partial \beta^2}\biggr)^{-1}\cdot\frac{\partial l}{\partial \beta}
\end{equation*}

If the absolute differences between $\beta_k$ and $\beta_{k+1}$ are below some predefined tolerance threshold, convergence can be declared and we decide we have obtained the estimated coefficients $\hat{\beta}$. Alternatively if the algorithm fails to converge, we set a maximum number number of iterations to prevent an infinite loop.

For our case, let us consider a single term within the the log-likelihood (no penalty):
\begin{equation*}
    l_i = y_i\log\theta_i + (1-y_i)\log(1-\theta_i)
\end{equation*}
Therefore:
\begin{equation*}
    \frac{\partial l_i}{\partial \theta_i} = \frac{y_i}{\theta_i}-\frac{1-y_i}{1-\theta_i}
\end{equation*}
and also:
\begin{equation*}
    \frac{\partial^2 l_i}{\partial \theta_i^2} = -\frac{y_i}{\theta_i^2}-\frac{1-y_i}{(1-\theta_i)^2}
\end{equation*}
which are both scalars. 

Recall that:
\begin{equation*}
    \theta_i = \frac{e^{d_i^T\beta}}{1+e^{d_i^T\beta}}
\end{equation*}
and so for $j,k = 1,2,...,p$, we have:
\begin{equation*}
    \frac{\partial\theta_i}{\partial\beta_j} = d_{ij}\frac{e^{d_i^T\beta}}{1+e^{d_i^T\beta}}-d_{ij}\frac{e^{d_i^T\beta}e^{d_i^T\beta}}{(1+e^{d_i^T\beta})^2} = d_{ij}\theta_i(1-\theta_i)
\end{equation*}
and:
\begin{equation*}
    \frac{\partial^2\theta_i}{\partial\beta_j\partial\beta_k} = d_{ij}d_{ik}(1-2\theta_i)\theta_i(1-\theta_i)
\end{equation*}

We can now derive the first and second derivatives of the likelihood function with respect to $\beta$ to be:
\begin{equation*}
    \frac{\partial l_i}{\partial\beta_j} = \frac{\partial l_i}{\partial\theta_i}\frac{\partial\theta_i}{\partial\beta_j}
\end{equation*}
and
\begin{equation*}
    \frac{\partial^2l_i}{\partial\beta_j\partial\beta_k} = \frac{\partial l_i}{\partial\theta_i}\frac{\partial^2\theta_i}{\partial\beta_j\partial\beta_k} + \frac{\partial^2l_i}{\partial\theta_i^2}\frac{\partial\theta_i}{\partial\beta_j}\frac{\partial\theta_i}{\partial\beta_k}
\end{equation*}

The Newton-Raphson method for finding $\hat{\beta}$ that maximises the log-likelihood for logistic regression with no additional penalty can therefore be expressed as:
\begin{equation*}
    \beta_{k+1} = \beta_k - \biggl(\frac{\partial l_i}{\partial\theta_i}\frac{\partial^2\theta_i}{\partial\beta_j\partial\beta_k} + \frac{\partial^2l_i}{\partial\theta_i^2}\frac{\partial\theta_i}{\partial\beta_j}\frac{\partial\theta_i}{\partial\beta_k}\biggr)^{-1}\biggl(\frac{\partial l_i}{\partial\theta_i}\frac{\partial\theta_i}{\partial\beta_j}\biggr)
\end{equation*}

Convergence problems can still exist using R's \texttt{glm.fit} when fitting a logistic regression model. This occurs when the number of parameters is large compared with the information in the data - as a result two different errors can occur, either the algorithm does not converge or it does converge but some of the estimated $\beta$ coefficients are very high. This is a common problem when utilising binary data as there is relatively little information that can be obtained at each single data point. There are two solutions that we can implement to avoid these errors: replicate the data several times (denoted \textit{nrep}) or reduce the number of parameters to be estimated by reducing the number of knots, $p_x$ and/or $p_z$. The selection of the number of replications and numbers of knots is not explored in this work, but as a result of these errors the default simulation set-up is $N_\mathcal{H}=400$ with two replicates of each observation, using $p_x=p_z=8$ knots when fitting a B-Spline estimate to achieve the design matrix $D$. 

\subsection{\textsc{Single Penalty - P-Spline Estimation}}
As was the case for the linear model we penalize using P-Spline estimations, selecting $\beta$ to maximise:
\begin{equation*}
    l(\beta) - \lambda_1\beta^T(P_1^TP_1+P_2^TP_2)\beta
\end{equation*}
in which $P_1$ and $P_2$ are the same row/column roughness matrices used previously in the linear model to prevent overfitting. A key difference from previous single penalty usage however is that we are now trying to maximise the objective likelihood function rather than minimise the least squares objective function - therefore the penalty is now subtracted rather than added. 

Defining this as the roughness penalty (RP):
\begin{equation*}
    RP = -\lambda_1\beta^T(P_1^TP_1+P_2^TP_2)\beta
\end{equation*}
we are able to find the first and second derivatives that are to be added to the terms we found in the previous chapter when using Newton-Raphson upon the no penalty method, such that:
\begin{equation*}
    \frac{\partial RP}{\partial\beta} = -2\lambda_1(P_1^TP_1+P_2^TP_2)\beta
\end{equation*}
and
\begin{equation*}
    \frac{\partial^2 RP}{\partial \beta^2} = -2\lambda_1(P_1^TP_1+P_2^TP_2)
\end{equation*}
Thus giving the overall first derivative term of the P-Spline Estimation using Newton-Raphson method:
\begin{equation*}
    \frac{\partial l_i}{\partial\beta_j} = \frac{\partial l_i}{\partial\theta_i}\frac{\partial\theta_i}{\partial\beta_j}-2\lambda_1\biggl[(P_1^TP_1+P_2^TP_2)\beta\biggr]_j
\end{equation*}
and second derivative:
\begin{equation*}
    \frac{\partial^2l_i}{\partial\beta_j\partial\beta_k} = \frac{\partial l_i}{\partial\theta_i}\frac{\partial^2\theta_i}{\partial\beta_j\partial\beta_k} + \frac{\partial^2l_i}{\partial\theta_i^2}\frac{\partial\theta_i}{\partial\beta_j}\frac{\partial\theta_i}{\partial\beta_k}-2\lambda_1\biggl[P_1^TP_1+P_2^TP_2\biggr]_{jk}
\end{equation*}
We note that the penalty term for the first derivative is a vector and a matrix for the second derivative, hence the $j$ and $j,k$ subscripts respectively. We use these values within our Newton-Raphson approximation as outlined previously to find the estimated $\hat{\beta}$ coefficients. 

\subsection{\textsc{Double Penalty - Marginal Penalization}}
We now wish to add the novel second penalty taking into account the discrepancies between the marginal estimated from the horizontal data $\mathcal{H}$ and vertical data $\mathcal{V}$. To reiterate, for simulation purposes we use the true marginal $\theta_{true}(x_{test})$ instead of the estimate $\hat{\theta}_\mathcal{V}(x_{test})$, noting that in reality this would not be possible but is helpful in simulations. 

As we did so in the linear model, we use a kernel smoothing method to estimate $\hat{\theta}_\mathcal{H}(x_0)$ from the fitted model $\hat{\theta}_\mathcal{H}(x_i,z_i)$, such that:
\begin{equation*}
    \hat{\theta}_\mathcal{H}(x_0) = \sum_{i \in \mathcal{H}}k\biggl(\frac{x_i-x_0}{\sigma_k}\biggr)\hat{\theta}_\mathcal{H}(x_i,z_i) \biggr/ \sum_{i \in \mathcal{H}}k\biggl(\frac{x_i-x_0}{\sigma_k}\biggr)
\end{equation*}
with $k(.)$ being a kernel function, $\sigma_k$ a smoothing parameter, $x_0$ being any element from $x_{test}$ (a scalar) and $(x_i,z_i)$ representing covariates for element $i$ within $\mathcal{H}$. We can express this in vector form:
\begin{equation*}
    \hat{\theta}_\mathcal{H}(x_0) = K\theta_\mathcal{H}(x,z)
\end{equation*}
where K is a $n_0 \times N_\mathcal{H}$ matrix of weights that have been suitably scaled. These weights do not contain $\beta$ and so therefore $K$ is a fixed constant in the optimisation of the maximum log-likelihood objective function. The objective function now contains two penalties, evaluated at test vector $x_0$. Note that for simulations we use $\theta(x_0)$ instead of $\hat{\theta}_\mathcal{V}(x_0)$ as mentioned previously. The objective function is as follows:
\begin{equation*}
    l(\beta)-\lambda_1\beta^T(P_1^TP_1+P_2^TP_2)\beta-\lambda_2\biggl(\hat{\theta}_\mathcal{H}(x_0)-\theta(x_0)\biggr)^T\biggl(\hat{\theta}_\mathcal{H}(x_0)-\theta(x_0)\biggr)
\end{equation*}

As before and for simplicity, we define the marginal penalty (MP) as:
\begin{equation*}
    MP = -\lambda_2\biggl(K\theta_\mathcal{H}(x,z)-\theta(x_0)\biggr)^T\biggl(K\theta_\mathcal{H}(x,z)-\theta(x_0)\biggr)
\end{equation*}
and for simplicity:
\begin{equation*}
\begin{split}
    MP &= -\lambda_2(K\theta-\theta_0)^T(K\theta-\theta_0) \\
    &= -\lambda_2\biggl({\theta^TK^TK\theta-2\theta_0^TK\theta+\theta_0^TK\theta+\theta_0^T\theta_0}\biggr) 
\end{split}
\end{equation*}

Only $\theta$ depends upon $\beta$. Differentiating the $i^{th}$ term of $\theta$ with respect to $\beta_j$:
\begin{equation*}
    \frac{\partial\theta_i}{\partial\beta_j}=d_{ij}\theta_i(1-\theta_i)
\end{equation*}
and then collecting these into an $N_\mathcal{H}$ vector $\partial\theta / \partial\beta_j$, we find the first derivatives of the marginal penalty as:
\begin{equation*}
    \frac{\partial MP}{\partial\beta_j} = -2\lambda_2(\theta^TK^TK-\theta^T_0K)\frac{\partial\theta}{\partial\beta_j}
\end{equation*}

Now differentiating the $i^{th}$ term of $\theta$ again, this time with respect to $\beta_k$:
\begin{equation*}
    \frac{\partial^2\theta_i}{\partial\beta_j\partial\beta_k} = d_{ij}d_{ik}(1-2\theta_i)\theta_i(1-\theta_i)
\end{equation*}
and then collecting these into an $N_\mathcal{H}$ vector, $\partial^2\theta / \partial\beta_j\beta_k$, we obtain the second derivatives of the marginal penalty as:
\begin{equation*}
    \frac{\partial^2MP}{\partial\beta_j\partial\beta_k} = -2\lambda_2\biggl\{\biggl(\frac{\partial\theta}{\partial\beta_k}\biggr)^TK^TK\frac{\partial\theta}{\partial\beta_j}+\theta^TK^TK\frac{\partial^2\theta}{\partial\beta_j\partial\beta_k}-\theta_0^TK\frac{\partial^2\theta}{\partial\beta_j\partial\beta_k}\biggr\}
\end{equation*}
We use these values within our Newton-Raphson approximation as outlined previously to find the estimated $\hat{\beta}$ coefficients.

\subsection{\textsc{Measure of Fit}}
As we have done previously, we will measure the fit of these three approaches using the sum of squares, comparing the estimated probabilities $\hat{\theta}_\mathcal{H}(x,z)$ with the true probabilities $\theta(x,z)$:

\begin{equation*}
    SS = \sum_{x,z}\biggl(\hat{\theta}_\mathcal{H}(x,z)-\theta(x,z)\biggr)^2
\end{equation*}

One difference in this approach however is we now take into account heterogeneous variances that may exist within $x$ and $z$, thus we use a weighted sum of squares as an alternative:
\begin{equation*}
    WSS = \sum_{x,z}\frac{\biggl(\hat{\theta}_\mathcal{H}(x,z)-\theta(x,z)\biggr)^2}{\theta(x,z)\biggl(1-\theta(x,z)\biggr)}
\end{equation*}

Varying sample size $N_\mathcal{H}$, number of knots $p_x=p_z$ and the relationship between covariates $x$ and $z$, displayed in Tables \ref{table:table5} and \ref{table:table6} are the mean average sum of squares for fitted and marginal values from 100 simulations of each dataset with these varying combinations. Note that each data point was replicated an appropriate amount of times in each dataset to prevent non-convergence errors, this is highlighted by 'nrep' in Tables \ref{table:table5} and \ref{table:table6}. The penalty parameters $\lambda_1$ and $\lambda_2$ are at their optimal values for each sample size, noise and covariate relationship combination, found through the method described in Section 5, and given in Table \ref{table:table7}.

\begin{table}[H]
\centering
\begin{adjustbox}{width=1\textwidth}
\begin{tabular}{cccc|ccc}
\textbf{Interaction} &  \textbf{\textbf{$N_\mathcal{H}$}} & \textbf{$p_x=p_z$} & \textbf{nrep} & \textbf{Fit0 WSS(Fitted)} & \textbf{Fit1 WSS(Fitted)} & \textbf{Fit2 WSS(Fitted)} \\ \hline
No & 100 & 4 & 4 & 16.76 & 11.86 & \textbf{9.22}\\
No & 100 & 8 & 4 & 16.50 & 10.34 & \textbf{8.66}\\
No & 400 & 8 & 2 & 17.75 & 12.85 & \textbf{9.26}\\
No & 400 & 18 & 2 & 37.99 & 12.32 & \textbf{9.13}\\
No & 900 & 8 & 1 & 17.94 & 12.62 & \textbf{9.14}\\
\hline
Yes & 100 & 8 & 8 & 83.63 & 32.44 & \textbf{30.79}\\
Yes & 400 & 8 & 2 & 74.55 & 22.98 & \textbf{20.71}\\
Yes & 900 & 8 & 1 & 73.20 & 22.48 & \textbf{20.18}\\
\end{tabular}
\end{adjustbox}
\caption{Mean average weighted sum of squares of fitted values for each model fit as model structure, sample size and number of knots are varied (100 simulations)}
\label{table:table5}
\end{table}

\begin{table}[H]
\centering
\begin{adjustbox}{width=1\textwidth}
\begin{tabular}{cccc|ccc}
\textbf{Interaction} &  \textbf{\textbf{$N_\mathcal{H}$}} & \textbf{$p_x=p_z$} & \textbf{nrep} & \textbf{Fit0 WSS(Marginal)} & \textbf{Fit1 WSS(Marginal)} & \textbf{Fit2 WSS(Marginal)} \\ \hline
No & 100 & 4 & 4 & 1.37 & 1.32 & \textbf{0.99}\\
No & 100 & 8 & 4 & 1.14 & 1.21 & \textbf{0.83}\\
No & 400 & 8 & 2 & 0.85 & 1.00 & \textbf{0.49}\\
No & 400 & 18 & 2 & 0.77 & 0.90 & \textbf{0.47}\\
No & 900 & 8 & 1 & 0.76 & 0.88 & \textbf{0.46}\\
\hline
Yes & 100 & 8 & 8 & 0.85 & 0.72 & \textbf{0.34}\\
Yes & 400 & 8 & 2 & 0.65 & 0.56 & \textbf{0.24}\\
Yes & 900 & 8 & 1 & 0.54 & 0.45 & \textbf{0.21}\\
\end{tabular}
\end{adjustbox}
\caption{Mean average weighted sum of squares of marginal values for each model fit as model structure, sample size and number of knots are varied (100 simulations)}
\label{table:table6}
\end{table}

\begin{table}[H]
\centering
\begin{adjustbox}{width=0.6\textwidth}
\begin{tabular}{cccc|ccc}
\textbf{Interaction} &  \boldmath{$N_\mathcal{H}$} & \boldmath{$p_x=p_z$} & \textbf{nrep} & \boldmath{$\lambda_{1a}$} & \boldmath{$\lambda_{1b}$} & \boldmath{$\lambda_{2}$} \\ \hline
No & 100 & 4 & 4 &  0.06 & 0.23 & 8.94\\
No & 100 & 8 & 4 &  0.21 & 0.22 & 8.92\\
No & 400 & 8 & 2 &  0.28 & 0.30 & 18.86\\
No & 400 & 18 & 2 &  6.34 & 7.06 & 18.98\\
No & 900 & 8 & 1 &  0.25 & 0.33 & 20.82\\

\hline
Yes & 100 & 8 & 8 &  0.71 & 0.67 & 13.06\\
Yes & 400 & 8 & 2 &  0.62 & 0.59 & 15.98\\
Yes & 900 & 8 & 1 &  0.57 & 0.59 & 18.46\\

\end{tabular}
\end{adjustbox}
\caption{Optimum penalty parameter values for each model structure and parameter combinations}
\label{table:table7}
\end{table}

We see from  Tables \ref{table:table5} and \ref{table:table6} above that the average sum of squares for the fitted and marginal values are all at their lowest when Fit2 which utilises the novel additional marginalisation penalty is applied. For datasets in which there is an interaction between covariates Fit1 always performs better than Fit0 when comparing the weighted sum of squares for the marginal values, however typically Fit0 outperforms Fit1 when there is no interaction between datasets. Fit1 however does perform better than Fit0 in all cases when comparing the sum of squares for fitted values. It is again encouraging to see that Fit2 works as expected and outperforms both single penalty and non-penalized methods. It is worth reiterating that comparing sums of squares between different parameter combinations is unwise, therefore comparisons should only be made between model fits for the same combinations. We also see in Table \ref{table:table7} that generally as $N_\mathcal{H}$ increases, the size of each penalty parameter increases also. For both interaction and non-interaction datasets the optimum values for $\lambda_1$ typically follow $\lambda_{1a}=\lambda_{1b}$, however the value of these penalty parameters greatly increases when the number of knots $p_x=p_z$ increases significantly to 18. Values for $\lambda_2$ also tend to increase as sample size increases also, albeit from a far greater initial value.

\section{\textsc{Application}}
\subsection{\textsc{Dataset}}
We have seen from our simulated datasets that the double penalty approach of a marginalisation penalty along with a P-Spline estimation can provide a better model fit for data with either a continuous response or a binary response. We now wish to test whether the double penalty method yields similar better fitting results upon our real data. The data we use within the chapter is a part of the European Non-Alcoholic Fatty Liver Disease (NAFLD) Registry \cite{hardy2020european}. The registry is made up of two subsections including a meta-cohort of all consented cases from European NAFLD studies from 2010-2017 and LITMUS (Liver Investigation: Testing Marker Utility in Steatohepatitis), an active study of more than 25 studies across 13 European countries that have been recruiting NAFLD cases since 1$^{st}$ January 2018. The data that is contained within the registry includes an individual's clinical information, liver histopathology, image data, quality of life questionnaires and biopsy samples amongst others. For a smaller sample of individuals, there is also multi-omics (genomic, epigenomic, transcriptomic, metabolomic, proteomic and metagenomic) data available, to help understand inter-patient variability in hepatic injury as well as the contribution of environmental factors to disease progression.

In previous works using the LITMUS dataset, 19 'core' covariates were chosen to predict nine binary responses of particular interest, which indicate whether a patient has reached a specific NAFLD stage that is considered to be serious and/or irreversible. These 19 covariates relate to standard measurements that are achieved through a blood test or a routine GP appointment, this includes information such as age, gender, BMI, pre-existing health conditions, as well as chemicals, proteins and cells in the blood. These covariates will form our vertical dataset $\mathcal{V}$. Focusing upon a binary response variable of 'At-Risk NASH', with positive case or '1' being defined as an individual having a NAFLD Activity Score (NAS) greater than or equal to 4 and a Fibrosis stage of greater than or equal to 2, and negative case '0' otherwise, this response variable is a key stage in the NAFLD natural progression between benign steatosis and more serious fibrosis and cirrhosis. There are approximately 6000 individuals that have an At-Risk NASH response and have had the 19 covariates that make up the vertical dataset $\mathcal{V}$ measured. The horizontal dataset $\mathcal{H}$ includes approximately 1500 individuals who have had the 19 core features of $\mathcal{V}$ measured, alongside multi-omics data, specifically genomic sequencing data. This includes 33 covariates of SNPs (single nucleotide polymorphisms) and four polygenic risk scores, ultimately resulting in 37 additional covariates within the horizontal dataset $\mathcal{H}$. 

A summary of the $\mathcal{V}$ and $\mathcal{H}$ datasets is illustrated below:

\begin{table}[H]
\centering
\begin{adjustbox}{width=0.4\textwidth}
\begin{tabular}{c|cc|cc}
\textbf{Dataset} &  \textbf{$N$} & \textbf{$p$} & \textbf{$Y=0$} & \textbf{$Y=1$} \\ \hline
$\mathcal{V}$ & 6024 & 19 & 4014 & 2010 \\
$\mathcal{H}$ & 1456 & 19+37 & 860 & 596 \\
\end{tabular}
\end{adjustbox}
\end{table}

\subsection{\textsc{Adaptations from Simulated Models}}
\subsubsection{\textsc{Dimensionality Reduction}}

As seen in Sections 4 and 6 our data simulations have been limited to working upon datasets where the number of covariates is equal to two. In principle the methods we have created would work for more than two covariates, but the number of parameters would become very large and the fits would therefore be unstable. Instead we will adopt three dimensionality reduction techniques to obtain the best linear combinations of the 19 and 37 covariates. These will be taken as $x$ and $z$ respectively. We discuss this issue of dimensionality reduction in the discussion part of this paper.

The first method of dimensionality reduction applied on the real data set is using the linear predictor fitted on the link scale following the fit of a generalised linear model (GLM) of the covariates to the response \cite{nelder1972generalized}. GLMs are formed using three components: a linear predictor - a linear combination of covariates and coefficients; a probability distribution - used to generate the response variable; and a link function - simply a function that 'links' together the linear predictors and probability distribution parameter. By fitting a GLM to $\mathcal{H}$ upon the 19 covariates that also exist within $\mathcal{V}$ with the corresponding binary response $y$ for these observations, we take the linear predictor fitted on the link scale and return a single vector that represents $x$. In the same way, we fit a GLM upon the 37 additional covariates that exist only within $\mathcal{H}$ with the corresponding binary response $y$ for these observations and take the linear predictor fitted on the link scale to return another single vector, this time representing $z$. This technique is common in prognostic modelling within medical domains where the linear predictor is often used as a prognostic index, i.e. a measure of future risk), for patients \cite{ramspek2021external, al2023bayes}.

The second method used in this section is principal component analysis (PCA). Developed by Karl Pearson \cite{pearson1901liii}, PCA is one of the most common methods of dimensionality reduction. In a nutshell, supposing we have $p$ covariates, PCA transforms $p$ variables $e_1,e_2,...,e_p$ called principal components, each of which are linear combinations of the original covariates $x_1,x_2,...,x_p$. We select coefficients for each covariate so that the first principal component $e_1$ explains the most variation within the data, and then the second principal component $e_2$ (uncorrelated with $e_1$) explains the next most variation, and so on. For our purpose we use the first principal component when performing a PCA on the 19 and then the 37 covariates, thus providing $x$ and $z$ as single vectors we can use within our analysis. 

The final method of dimensionality reduction we use is t-distributed stochastic neighbour embedding (tSNE). Based upon the van der Maaten t-distributed variant of stochastic neighbour embedding, developed by Hinton and Roweis \cite{van2008visualizing} \cite{hinton2002stochastic}, tSNE, unlike the linear predictor and PCA methods, is a non-linear technique that aims to preserve pairwise similarities between data points in a low-dimensional space. The tSNE method calculates the pairwise similarity of data points within high and low dimensional space and assigns high and low probabilities to data points that are are close and far away from a selected data point respectively. It then maps the higher dimensional data onto a lower dimensional space whilst minimizing the divergence in the probability distributions of data points within the high and lower dimensional data. This mapping then provides a vector which can be used within our methods, to represent both $x$ and $z$ variables. The greatest difference between the PCA and tSNE methods are that PCA aims to preserve the variance of the data whereas tSNE aims to preserve the relationship between the data points. 

\subsubsection{\textsc{Estimating \boldmath{$\hat{\theta}_\mathcal{V}(x)$}}}
Recall that for a binary response variable $y$
\begin{equation*}
    Pr(Y = 1 | x,z) = \theta(x,z)
\end{equation*}
where $\theta(x,z)$ is a smooth but unknown function of all fitted probabilities. The associated marginal is given:
\begin{equation*}
    Pr(Y = 1 | x) = \theta(x) = \int_z \theta(x,z) f_Z(z | x) dz
\end{equation*}
in which $f_Z(z | x)$ is the conditional probability density function of $z$ given $x$.

As we have done previously we are going to estimate $\theta(x,z)$ from $\mathcal{H}$ but modify our estimate to make sure that $\hat{\theta}_\mathcal{H}(x)$, i.e. the estimated marginal $x$ attained from our horizontal data, is close to the more accurate marginal from our vertical data, $\hat{\theta}_\mathcal{V}(x)$. One change from the simulations that were carried out is that previously we assumed that $\mathcal{V}$ was very large so the uncertainty surrounding $\hat{\theta}_\mathcal{V}(x)$ was very small - in simulations we therefore used the true marginal $\theta(x)$. This however is not possible to calculate in real applications data, so our first task is therefore to estimate $\hat{\theta}_\mathcal{V}(x)$. 

As shown in Section 6, we can use a P-Spline approach using a logistic model with a smooth predictor. Using a B-Spline approximation and a design matrix $D$ for the vertical data $\mathcal{V}$, for any case $i \in \mathcal{V}$:
\begin{equation*}
    \theta(x_i) = \frac{e^{d_i^T\beta}}{1+e^{d_i^T\beta}}
\end{equation*}
in which $d_i^T$ is row $i$ of $D$. We then penalize as previous using the penalty coefficient $\lambda_1$.

\subsubsection{\textsc{Marginal in the Second Penalty}}
In our simulations to use the second marginalisation penalty, we compare the marginal from our fit using $x$ and $z$ in our horizontal data, and compare this with the true marginal values. Naturally the true marginal is unknown in our real data, so we therefore use an accurate estimate from $\mathcal{V}$. As mentioned in Section 3, we use a predefined vector $x_{test}$ to calculate the marginal from $\mathcal{H}$. In our real data we can now simply use $x_\mathcal{H}$, the observed $x$-values from $\mathcal{H}$ to calculate the marginal. The advantage of this is that we use all values in $x$ from $\mathcal{H}$ rather than just unique values, allowing the second penalty term to have greater weight for more common values of $x$. Another advantage is the marginal from $\mathcal{H}$, $\hat{\theta}_\mathcal{H}(x_i)$ for $i \in \mathcal{H}$ is produced as a part of the fitting procedure. 

\subsubsection{\textsc{Means of Comparison}}
Within simulations, we were able to know with certainty the true smooth function $\theta(x,z)$, and the true marginal of $x$, $\theta(x)$. This meant that we were able to compare the fit of models using the sum of squares of fitted values by comparing $\theta(x,z)$ with the model fits $\hat{\theta}_\mathcal{H}(x,z)$, and also through the sum of squares of the marginal values by comparing estimated marginal $\hat{\theta}_\mathcal{H}(x)$ with the true marginal $\theta(x)$. 

In our real data as mentioned above we now do not have perfect knowledge and do not know $\theta(x,z)$, therefore a comparison by means of sum of squares of fitted values is now redundant. This is because it is not possible to estimate from $\hat{\theta}_\mathcal{V}(x,z)$ as $z$ does not exist, and $\theta_\mathcal{H}(x,z)$ is unknown. Allowing the estimated marginal of $x$ from $\mathcal{V}$ to now replace the true marginal $\theta(x)$, our only means of comparison now is through sum of squares of the marginal values, with values closer to zero indicating a greater fit. In the discussion section of this paper we mention possible future work of evaluating model fits upon data in which we do not have perfect knowledge. 

\subsubsection{\textsc{Cross-Validation for Choosing Penalty Parameters}}
One final adaptation from modelling upon real data to simulations is that it is now feasible to undertake a $k$-fold cross-validation method in order to determine smoothing parameter $\lambda_1$. Recall in simulations $\lambda_1$ was selected through comparing the sum of squares values when fitting the simulated data across a grid of $\lambda_1$ values. This sum of squares value was found through comparing model fit values to the truth therefore cross-validation was not necessary, it would have also taken a long time to carry out due to the number of simulations and parameter combinations. In our real data application, k-fold cross-validation \cite{larson1931shrinkage} is now required as ground truth is unknown. 

Setting number of folds $k=10$ and allowing for a 90:10 train/test split, each train set data is fitted using a P-Spline approximation whilst iterating through a grid of $\lambda_1$ values. The coefficients of each of these fits are then multiplied with the design matrix created from the test set and then put into an expit function to give the estimated fitted probabilities $\hat{\theta}_{test}(x,z)$. We use three different metrics to compare $\hat{\theta}_{test}(x,z)$ and the values of $y$ within the test set: sum of squares (SS), log-likelihood (LL) and area under curve (AUC). For each $\lambda_1$ value, the median value for each metric across the $k=10$ folds is found. The 'best' $\lambda_1$ value is therefore the median value that is either the smallest SS, or greatest LL or AUC value. 

We then use all three supposed 'best' $\lambda_1$ values according to these metrics to find $\lambda_2$. This is simply found through using these $\lambda_1$ values and scanning through a grid of $\lambda_2$ values, until an acceptable improvement in fit from using the additional marginalisation penalty over the single penalty P-Spline approximation is found, in our case this acceptable improvement is 50\% reduction in sum of squares in marginal values. We can also select $\lambda_2$ as simply the value that provides the lowest sum of squares of marginal values when using the additional marginalisation penalty.

We accept that our method of selecting $\lambda_2$ is ad-hoc and discuss potential future work options to select $\lambda_2$ values in the discussion. Increasing $\lambda_2$ values will take $\hat{\theta}_\mathcal{H}(x_{test})$ values ever closer to $\hat{\theta}_\mathcal{V}(x_{test})$ at the expense of a poorer, more biased estimate of $\theta(x,z)$. In practice we would like $\hat{\theta}_\mathcal{H}(x_{test})$ to be just close enough to $\hat{\theta}_\mathcal{V}(x_{test})$ to consider a realistic and feasible estimate of the underlying true $\theta(x_{test})$. This depends upon the level of noise that exists in $\hat{\theta}_\mathcal{H}(x_{test})$ and to a lesser extent the noise in $\hat{\theta}_\mathcal{V}(x_{test})$. Selecting $\lambda_2$ based upon a 50\% reduction in SS of marginal values is therefore preferable rather than the outright best SS value - this is because we accept that there is a level of noise in $\hat{\theta}_\mathcal{H}(x_{test})$ and that it would not be exactly the same as $\theta(x_{test})$ even if we had perfect knowledge on the correct marginal, we just expect these values to be close. By increasing $\lambda_2$ we force these values to be closer together, leading to more bias within $\hat{\theta}_\mathcal{H}(x,z)$.

\subsection{\textsc{Results}}
Following the dimensionality reduction of the real dataset, $x$ and $z$ are now single vectors. In Figure \ref{fig:fig16} are nine scatter plots of the three $x$ and three $z$ values of the real data after being reduced via the three methods outlined in Section 7.2.1:

\begin{figure}[H]
    \centering
    \includegraphics[width=1.0\textwidth]{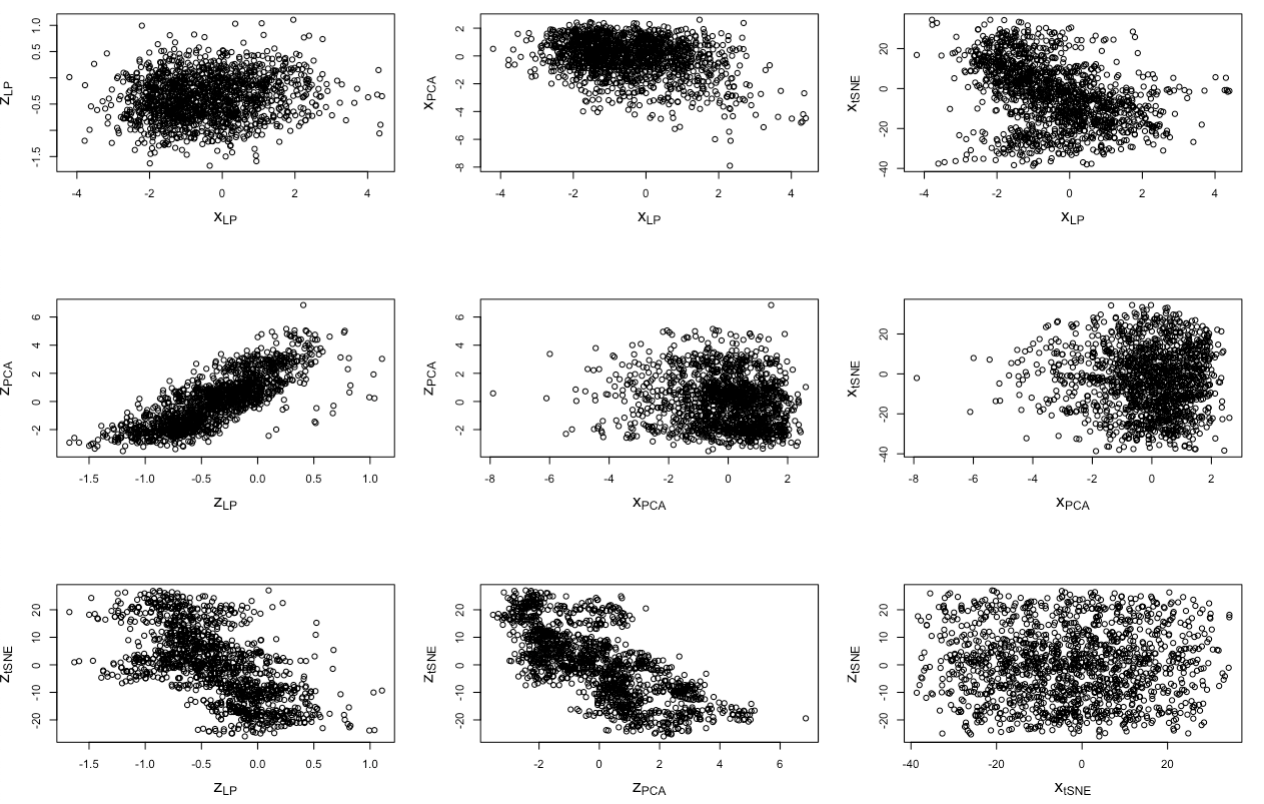}
    \caption{Scatter plot of each single $x$ and $z$ vector achieved following dimensionality reduction via Linear Predictor, PCA and tSNE. Note the subscript on each x and y axis label to highlight which $x$ and $z$ vector is being plotted.}
    \label{fig:fig16}
\end{figure}

 Errors frequently arose at two points during the modelling process for the application data. The first occasion is when fitting the three model types upon the newly scaled data, and the second occasion is when cross-validating upon the data to find optimal values for $\lambda_1$. In the second instance, errors occur in particular for values of high $\lambda_1$. For both occasions, this is due to the algorithm for fitting a generalised linear model either not converging or producing coefficient $\beta$'s that are ridiculously high. This error occurs when the fitted probabilities are extremely close to 0 or 1, this occurs when the predictor variable $x$ is able to perfectly separate the response variable. The consequence of this is that maximum likelihood estimates of the coefficients do not exist, and therefore the algorithm fails to converge. These errors can be alleviated by trimming the scaled values for $x$ and $z$ by removing extreme values at either end of the range.

 Following an extensive search of altering the minimum and maximum values of $x$ and $z$, the minimum number of data points that are removed without causing either a fitting or cross-validation error was 24 for both the interaction and non-interaction dataset when using PCA as a form of dimensionality reduction. This compares with 46 data points removed in the non-interaction dataset and 52 data points removed in the interaction dataset when using the Linear Predictor as a means of dimensionality reduction, and no data points removed for both the non-interaction and interaction datasets when using tSNE as the means of dimensionality reduction. 

Table \ref{table:table8} lists all results of the modelling upon the real applications data set. The measure of fit for the three fitted models is
 the sum of squares of the marginal values, attained from the horizontal data $\mathcal{H}$ and estimated vertical data $\mathcal{V}$ marginals. The method of dimensionality reduction, whether there is an interaction between $x$ and $z$, and the method of determination for $\lambda_1$ and $\lambda_2$ are all varied within these results.

\begin{landscape}
\begin{table}[htbp]
\centering
\begin{adjustbox}{width=1.35\textwidth}
\begin{tabular}{c|c|cc|cc|ccc}
\textbf{Dim. Reduction} & \textbf{Interaction} & \textbf{$\lambda_1$ Determination} & \textbf{$\lambda_2$ Determination} & \textbf{$\lambda_1$} & \textbf{$\lambda_2$} & \textbf{Fit 0 SS(Marg)} & \textbf{Fit 1 SS(Marg)} & \textbf{Fit 2 SS(Marg)} \\ \hline
Linear Predictor & No & SS & 50\% Improvement & 4 & 1.2 & 23.80 & 22.94 & \textbf{11.32} \\ 
Linear Predictor & No & SS & Best Fit2 & 4 & 2.2 & 23.80 & 22.94 & \textbf{7.66} \\
Linear Predictor & No & Log-likelihood & 50\% Improvement & 3 & 1.2 & 23.80 & 22.80 & \textbf{11.12} \\
Linear Predictor & No & Log-likelihood & Best Fit2 & 3 & 2.2 & 23.80 & 22.80 & \textbf{7.39} \\
Linear Predictor & No & AUC & 50\% Improvement & 6 & 1.3 & 23.80 & 23.16 & \textbf{11.11} \\
Linear Predictor & No & AUC & Best Fit2 & 6 & 2.3 & 23.80 & 23.16 & \textbf{7.70} \\ \hline

Linear Predictor & Yes & SS & 50\% Improvement & 4 & NA & - & - & - \\ 
Linear Predictor & Yes & SS & Best Fit2 & 4 & 2.4 & 24.69 & 25.83 & \textbf{13.42} \\
Linear Predictor & Yes & Log-likelihood & 50\% Improvement & 3 & NA & - & - & - \\
Linear Predictor & Yes & Log-likelihood & Best Fit2 & 3 & 2.4 & 24.69 & 25.66 & \textbf{13.09} \\
Linear Predictor & Yes & AUC & 50\% Improvement & 6 & NA & - & - & - \\
Linear Predictor & Yes & AUC & Best Fit2 & 6 & 2.4 & 24.69 & 26.03 & \textbf{13.76} \\  \hline

PCA & No & SS & 50\% Improvement & 2 & 1.0 & 42.37 & 33.85 & \textbf{16.30} \\ 
PCA & No & SS & Best Fit2 & 2 & 2.0 & 42.37 & 33.85 & \textbf{9.81} \\
PCA & No & Log-likelihood & 50\% Improvement & 2 & 1.0 & 42.37 & 33.85 & \textbf{16.30} \\
PCA & No & Log-likelihood & Best Fit2 & 2 & 2.0 & 42.37 & 33.85 & \textbf{9.81} \\
PCA & No & AUC & 50\% Improvement & 2 & 1.0 & 42.37 & 33.85 & \textbf{16.30} \\
PCA & No & AUC & Best Fit2 & 2 & 2.0 & 42.37 & 33.85 & \textbf{9.81} \\ \hline

PCA & Yes & SS & 50\% Improvement & 3 & 1.1 & 36.20 & 35.07 & \textbf{17.49} \\ 
PCA & Yes & SS & Best Fit2 & 3 & 2.0 & 36.20 & 35.07 & \textbf{12.12} \\
PCA & Yes & Log-likelihood & 50\% Improvement & 2 & 1.1 & 36.20 & 34.85 & \textbf{17.28} \\
PCA & Yes & Log-likelihood & Best Fit2 & 2 & 2.0 & 36.20 & 34.85 & \textbf{11.87} \\
PCA & Yes & AUC & 50\% Improvement & 1 & 1.1 & 36.20 & 34.55 & \textbf{16.91} \\
PCA & Yes & AUC & Best Fit2 & 1 & 2.0 & 36.20 & 34.55 & \textbf{11.48} \\ \hline

tSNE & No & SS & 50\% Improvement & 3 & 1.4 & 46.66 & 36.29 & \textbf{17.54} \\ 
tSNE & No & SS & Best Fit2 & 3 & 2.0 & 46.66 & 36.29 & \textbf{14.24} \\
tSNE & No & Log-likelihood & 50\% Improvement & 3 & 1.4 & 46.66 & 36.29 & \textbf{17.54}  \\
tSNE & No & Log-likelihood & Best Fit2 & 3 & 2.0 & 46.66 & 36.29 & \textbf{14.24} \\
tSNE & No & AUC & 50\% Improvement & 8 & 1.5 & 46.66 & 34.25 & \textbf{16.74} \\
tSNE & No & AUC & Best Fit2 & 8 & 2.0 & 46.66 & 34.35 & \textbf{14.26} \\ \hline

tSNE & Yes & SS & 50\% Improvement & 5 & NA & - & - & - \\ 
tSNE & Yes & SS & Best Fit2 & 5 & 2.0 & 48.59 & 53.35 & \textbf{30.19} \\
tSNE & Yes & Log-likelihood & 50\% Improvement & 5 & NA & - & - & - \\ 
tSNE & Yes & Log-likelihood & Best Fit2 & 5 & 2.0 & 48.59 & 53.35 & \textbf{30.19} \\
tSNE & Yes & AUC & 50\% Improvement & 0 & 2.1 & 48.59 & 48.59 & \textbf{21.00} \\
tSNE & Yes & AUC & Best Fit2 & 0 & 2.1 & 48.59 & 48.59 & \textbf{21.00} \\
\end{tabular}
\end{adjustbox}
\caption{Complete results for modelling upon NAFLD dataset}
\label{table:table8}
\end{table}
\end{landscape}

From Table \ref{table:table8} we can report that for every combination of dimensionality reduction, choice of $\lambda_1$ and $\lambda_2$, interaction or no interaction between covariates, Fit2 always provides a notable enhancement in marginal fit, this is because after all setting $\lambda_2=0$ would at worst reduce Fit2 to Fit1. Generally for interaction datasets, the additional penalty model offers less of an improvement in comparison to non-interaction datasets - in some cases such as the modelling upon an interaction dataset using linear predictor as a dimensionality reduction method and using AUC as the $\lambda_1$ determination method, there is no value of $\lambda_2$ that offers a greater than 50\% improvement in Fit2 over Fit1. This is the case for four other instances as shown in Table \ref{table:table7}. This is arbitrary however as it is clear from our results that Fit2 always provides an improvement in fit compared to Fit1 and Fit0. It is also notable that the P-Spline approximation method does not always offer an improvement upon the standard linear model. 

Generally log-likelihood and sum of squares methods select the same values for $\lambda_1$, however there is more variation when AUC is the method of determination for $\lambda_1$. Values for $\lambda_2$ are almost always identical regardless of $\lambda_1$ determination method, typically offering $\lambda_2$ approximately equal to 2 when selecting Fit2 purely on best fit, and $\lambda_2$ approximately equal to 1 when selecting $\lambda_2$ based upon a 50\% improvement in fit for Fit2 over Fit1. 

In Figure \ref{fig:figure17} we graphically compare the marginal estimates from $\mathcal{H}$, $\hat{\theta}_\mathcal{H}(x)$, which we receive from each model fit, this with our estimated marginal of $x$ from $\mathcal{V}$ for each model fit and each dimensionality reduction method. The red lines in each plot represent $\hat{\theta}_\mathcal{V}(x)$ and blue lines represent $\hat{\theta}_\mathcal{H}(x)$. The top row of plots are obtained through using the Linear Predictor as a means of dimensionality reduction; the middle row via PCA; and bottom row through tSNE. Each graph on the left of the plot illustrates the fitted marginal probabilities achieved from Fit0; the middle via Fit1; and the right hand side via Fit2. Better model fit is demonstrated the closer the blue and red lines are to one another, and as we can see for Fit2 $\hat{\theta}_\mathcal{H}(x)$ (red) is closest to $\hat{\theta}_\mathcal{V}(x)$ (blue) for all dimensionality reduction methods. Fit2 is therefore a better fit for our applications data compared to Fit0 and Fit1. 

\begin{landscape} \centering
\begin{figure}[H]
    \centering
    \includegraphics[width=1.3\textwidth]{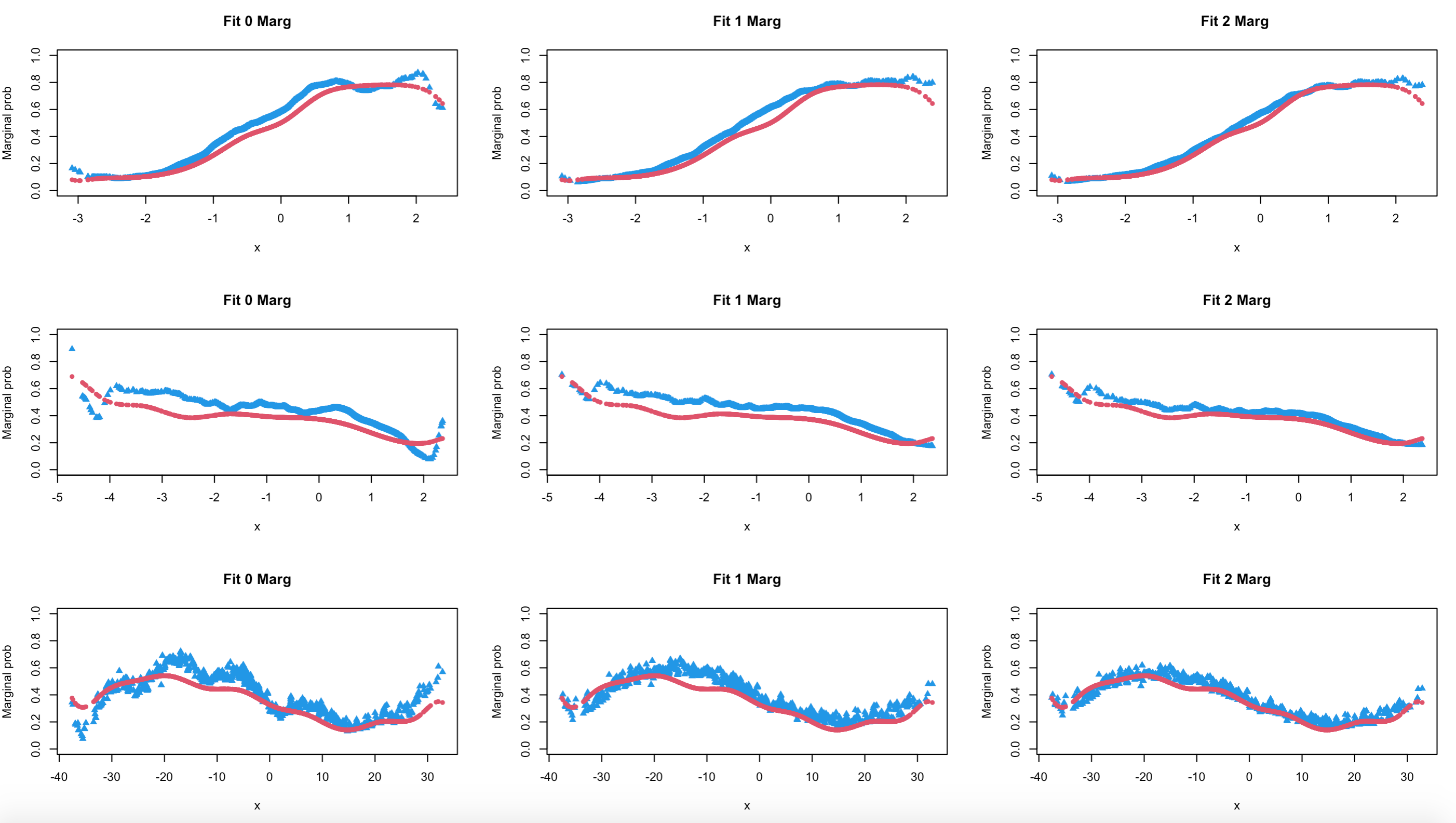}
    \caption{Comparison between $\hat{\theta}_\mathcal{V} (x)$ (Red) and $\hat{\theta}_\mathcal{H}(x)$ (Blue) for each model fit and each dimensionality reduction: Linear Predictor (top row) / PCA (middle row) / tSNE (bottom row)}.
    \label{fig:figure17}
\end{figure}
\vspace{10mm}
\end{landscape}

\section{\textsc{Discussion and Future Work}}
To our knowledge this is the first work to propose additional marginal penalties in a flexible regression. There are however a number of areas for future development. The first is that we were unable to develop a succinct method of selecting penalty parameter $\lambda_2$, relating to the discrepancies between marginal values of $x$; we relied upon cross-validation to select $\lambda_1$ however this method is not possible in selection for $\lambda_2$ - we therefore relied upon the manual scanning across a range of values to select $\lambda_2$. We are prepared to accept a slightly worse fit to the data in $\mathcal{H}$ if a more realistic marginal when compared with that from $\mathcal{V}$ is obtained. This means we are we are not trying to optimise the fit but we desire as good a fit as possible subject to the marginal estimate $\hat{\theta}_\mathcal{H}(x)$ being consistent with $\hat{\theta}_\mathcal{V}(x)$. Future work therefore would include the development of a concise method to choose $\lambda_2$. We know that as $\lambda_2 \to \infty$, $\hat{\theta}_\mathcal{H}(x) \to \hat{\theta}_\mathcal{V}(x)$, therefore one possibility for future work would be to gradually increase $\lambda_2$ until a consistent estimator is reached. Another method would include a computationally intense method of iterating through a grid of $\lambda_2$ values again, this time fitting our model to a sample of values within $\mathcal{H}$ for each $\lambda_2$ value. For each iteration, we can receive the marginal values $\hat{\theta}_\mathcal{H}(x)$ from these samples as well as a percentage confidence band for $\hat{\theta}_\mathcal{H}(x)$. If $\hat{\theta}_\mathcal{V}(x)$ lies within this band also, then we accept this $\lambda_2$ value as the 'optimal', and if not we try the next value in our defined $\lambda_2$ grid. An alternative solution could be to withhold some of $\mathcal{H}$ to assess fit, through dividing $\mathcal{H}$ into parts similar to a train/test split, using the train set to determine penalty parameter values $\lambda_1$ and $\lambda_2$ and using the test set to evaluate model performance with these selected values. 

Secondly, we are reliant upon the sum of squares of the marginal values to be our sole measurement of fit for modelling within our application to real data section. As seen in simulations we also used the sum of squares of fitted values as a means for comparison between different model fits, however with the true fitted values now unknown the method we utilised within simulations is now not feasible. Future work would therefore include the development of other methods of evaluating performance of our model when ground truth is unknown. One possible method of doing so would be to use two thirds of $\mathcal{H}$ to cross-validate and calculate optimal penalty parameter values $\lambda_1$ and $\lambda_2$ and using the remaining third of $\mathcal{H}$ to assess fit. This however is not entirely necessary as our aim for this work was not to model $\mathcal{H}$ as well as possible but to integrate the smaller cohort $\mathcal{H}$ and larger cohort $\mathcal{V}$ into a predictive analysis without the requirement of imputation. We would therefore consider accepting a slightly worse model fit for $\mathcal{H}$ to ensure a marginal that is closer to $\mathcal{V}$. 

Furthermore, we have developed our method for the case of $x$ and $z$ being scalars in Section 7.2.1. We used dimensionality reduction methods to reduce multivariate covariates to scalar summaries. These techniques naturally come with their own disadvantages. For example, if data is strongly non-linear dimensionality techniques such as PCA can struggle to fully capture covariate relationships, potentially resulting in a loss of information. Dimensionality reduction can also result in difficult to interpret transformations of covariates and are not always easy to visualize. Future work therefore includes being able to develop our methods and additional marginalisation penalty to work upon datasets without the need of transforming $\mathcal{H}$ covariates into single $x$ and $z$ vectors. Finally we have only considered non-parametric models for representing both $\mathcal{V}$ and $\mathcal{H}$. We mentioned in Section 2 our motivation and the suitability for non-parametric modelling, noting that if $f(y|x,z)$ takes a parametric modelling form it is unlikely $f(y|x)$ could be of the same parametric form also. It is however possible for one of $\mathcal{V}$ or $\mathcal{H}$ to be modelled parametrically provided the other is modelled non-parametrically. This is therefore another potential avenue for future work. 

\section{\textsc{Conclusion}}
Referring back to the purpose of this research, it is a common issue within data science of how to maximise the level of information that can be attained from asymmetric overlapping datasets. In a medical context, we have highlighted how particular subjects may have more information available to utilise within predictive analysis than the more common baseline information, such as specialist testing. Common solutions to this problem involve missing data imputation or simply two separate predictive models, one using baseline information only on a large number of individuals and one using baseline plus specialist testing information on a select number of individuals. The issue with missing data imputation is it is infeasible and bad practise to impute large levels of missing data, particularly if the cohort with larger levels of information available is substantially smaller than that of the larger cohort with less information. Utilising two separate predictive models for each cohort limits analysis and what we can learn from both the response variable and its interaction with covariates.

In this paper we propose a method to integrate the smaller cohort, named horizontal data ($\mathcal{H}$) and the larger cohort, named vertical data ($\mathcal{V}$) without the requirement for data imputation or data deletion. Simplifying the number of covariates down to two, $x$ and $z$, in which $x$ represents covariates every individual has recorded, and $z$ represents the added covariates only individuals within $\mathcal{H}$ have recorded, we are motivated by non-parametric models for modelling each cohort. We find that utilising flexible smoothing via B-Splines offers opportunities to take into account both cohorts into our analysis. Flexible smoothing models provide more robustness and flexibility to model complexly distributed data points where linear and polynomial regression models are unsatisfactory. Smoothness can be controlled by the introduction a penalty term to B-Splines, also known as a P-Spline - these penalties are desirable to prevent over/under-fitting to data. By looking at discrepancies between the marginal value of $x$ obtained from $\mathcal{H}$, denoted $\hat{\theta}_\mathcal{H}(x)$ with the marginal value of $x$ obtained from $\mathcal{V}$, denoted $\hat{\theta}_\mathcal{V}(x)$, we introduce a second penalty term to be able to model $\mathcal{H}$ whilst taking into account $\mathcal{V}$.

Through a series of data simulations, penalty parameter tunings and model adaptations to take into account both a continuous and binary response, we found that the model with the additional marginalisation penalty appended to a P-Spline approximation method outperformed both the linear B-Spline method and the standard P-Spline approximation method utilising the single smoothing penalty. Applying the model to a real life healthcare dataset with binary response relating to an individual's risk of developing Non-Alcoholic Steatohepatitis (NASH), we let $\mathcal{V}$ represent individuals who had a routine blood test taken, and $\mathcal{H}$ represent individuals who had further specialist, genomic sequencing data collected. We found similar results in that this model with the additional marginalisation penalty fitted the marginal values of the data better than both the linear B-Spline model and single penalty P-Spline approximation. 

Areas for future work include the development of a succinct method to select penalty parameter $\lambda_2$ and the finding of a measurement to take into account overall model fit when applying models to a real world dataset. In this work we omitted this, as our overall aim was to develop a method in which we could integrate asymmetric datasets into a predictive analysis upon a binary target, and therefore we had less of a focus on model fit. Future work will also include adapting our method to not require dimensionality reduction and also to consider parametric modelling for one of the $\mathcal{V}$ and $\mathcal{H}$ datasets. We have shown in this paper that the novel additional marginalisation penalty improved the fit of models as opposed to standard B-Spline and P-Splines approximation methods. These results are encouraging and illustrate a novel technique of how it is possible to integrate asymmetric datasets that share common levels of information without the need for data imputation or separate predictive modelling.

% \begin{comment}
%references 
\fancyhf{} %sets all head and foot elements empty
\fancyfoot[C]{\thepage}
% \section*{References}
% \addcontentsline{toc}{section}{References}
% \renewcommand*{\refname}{Reference}
% \vspace{-12mm}
\urlstyle{same}
{\begingroup
\footnotesize
\RaggedRight
\begin{spacing}{0.5} %0.95
    \bibliographystyle{ieeetr} %unsrturl "url" adds url in references list
    \bibliography{references}
\end{spacing}
\endgroup}

\newpage
\section*{\textsc{Appendix}}

\subsection*{\textsc{Construction of Design Matrices}}
As outlined in Section 3.1 there are two relationships response $y$ has with covariates $(x,z)$. The first instance of the smoothing relationship relating to there being no interaction between response $y$ and covariates $(x,z)$, suggests that given a predefined number of knots $p_x$, a B-Spline basis is fitted to covariate $x$ to provide the B-Spline basis matrix $B_x$, which has dimensions $N_\mathcal{H} \times p_x$, i.e. the number of observations within the $\mathcal{H}$ dataset by the number of knots $p_x$. This matrix represents the list of basis functions across all predefined knots, evaluated at each observation within $N_\mathcal{H}$. Similarly, fitting a B-Spline basis function to covariate $z$ with predefined number of knots, $p_z$, basis matrix $B_z$ with dimensions $N_\mathcal{H} \times p_z$ is outputted. The design matrix $D$ is then constructed by appending $B_z$ to $B_x$, and then adding an intercept term. The design matrix therefore has dimensions $N_\mathcal{H} \times (1+p_x+p_z)$.

For the second instance of the smoothing relationships, in this case where response $y$ has an interaction with covariates $(x,z)$, the design matrix $D$ is constructed differently. Matrices $B_x$ and $B_z$ are both constructed in the same way as before, however $D$ is now achieved through taking all products of a column in $B_x$ and a column in $B_z$ and then adding an intercept term. This therefore provides the design matrix $D$ with the dimensions $N_\mathcal{H} \times (1+p_xp_z)$.

\subsection*{\textsc{Construction of Roughness Matrices}}
Section 3.3 introduces P-Spline estimation as a means of penalizing the B-Spline, done so through the creation of penalty roughness matrices $P_1$ and $P_2$. The way $P_1$ and $P_2$ are constructed depends upon the relationship between response $y$ and covariates $(x,z)$. When there is an interaction, $P_1$ is found through the product between the identity matrix, $I$, of dimensions $p_x \times p_x$ and the difference matrix of dimensions $(p_x-2) \times p_x$, plus an intercept term, thus giving $P_1$ the dimensions of $(p_x(p_x-2)+1) \times p_xp_x$. Similarly, $P_2$ is found in the exact same way, using number of splines $p_z$ this time. $P_2$ therefore has the dimensions of $(p_z(p_z-2)+1) \times p_zp_z$.

When there is no interaction between the response and covariates, roughness matrices $P_1$ and $P_2$ have identical dimensions. In this case $P_1$ and $P_2$ take the dimensions of $[(p_x-2)+(p_z-2)+1 \times (p_z)+(p_z)+1]$, this is simply the two difference matrices applied to covariates $x$ and $z$ appended together, with an added intercept term. 
% \end{comment}

\end{document}